\documentclass[prd,eqsecnum,twocolumn,amsfonts,amssymb]{revtex4}

\usepackage{CJK}

\usepackage{graphicx}

\usepackage{bm}

\newcommand{\beq}{\begin{equation}}
\newcommand{\eeq}{\end{equation}}
\newcommand{\beqs}{\begin{eqnarray}}
\newcommand{\eeqs}{\end{eqnarray}}
\newcommand{\lsim}{\mathrel{\raisebox{-
.6ex}{$\stackrel{\textstyle<}{\sim}$}}}
\newcommand{\gsim}{\mathrel{\raisebox{-
.6ex}{$\stackrel{\textstyle>}{\sim}$}}}

\newcommand{\drawsquare}[2]{\hbox{%
\rule{#2pt}{#1pt}\hskip-#2pt
\rule{#1pt}{#2pt}\hskip-#1pt
\rule[#1pt]{#1pt}{#2pt}}\rule[#1pt]{#2pt}{#2pt}\hskip-#2pt
\rule{#2pt}{#1pt}}
\newcommand{\fund}{\raisebox{-.5pt}{\drawsquare{6.5}{0.4}}}
\newcommand{\sym}{\raisebox{-.5pt}{\drawsquare{6.5}{0.4}}\hskip-0.4pt%
        \raisebox{-.5pt}{\drawsquare{6.5}{0.4}}}
\newcommand{\asym}{\raisebox{-3.5pt}{\drawsquare{6.5}{0.4}}\hskip-6.9pt%
        \raisebox{3pt}{\drawsquare{6.5}{0.4}}}

\begin{document}

\begin{CJK*}{UTF8}{}

\title{Renormalization-Group Evolution of Chiral Gauge Theories}

\author{Yan-Liang Shi (\CJKfamily{bsmi}石炎亮) and Robert Shrock} 

\affiliation{C. N. Yang Institute for Theoretical Physics, 
Stony Brook University, Stony Brook, N. Y. 11794 }

\begin{abstract}

We calculate the ultraviolet to infrared evolution and analyze possible types
of infrared behavior for several asymptotically free chiral gauge theories with
gauge group SU($N$) and massless chiral fermions transforming according to a
symmetric rank-2 tensor representation $S$ and $N+4$ copies (flavors) of
a conjugate fundamental representation $\bar F$, together with a vectorlike
subsector with chiral fermions in higher-dimensional representation(s). We
construct and study three such chiral gauge theories. These have respective 
vectorlike subsectors comprised of (a) $p$ copies of fermions in the adjoint
representation, (b) $N = 2k$ even and $p$ copies of fermions in the
antisymmetric rank-$k$ tensor representation, and (c) $p$ copies of $\{S + \bar
S\}$ fermions.  Results are presented for beta functions, their infrared zeros,
and predictions from the most-attractive-channel approach for the formation of
bilinear fermion condensates.  Importantly, we show that for these theories,
the expected ultraviolet to infrared evolution obeys a conjectured inequality
concerning the field degrees of freedom for all values of the parameters $N$
and $p$ characterizing each theory.

\end{abstract}


\maketitle

\end{CJK*}


\section{Introduction}
\label{intro}

The question of how the properties of an asymptotically free chiral gauge
theory change as a function of the Euclidean momentum scale $\mu$ at which one
measures these properties is of fundamental physical interest. For sufficiently
large $\mu$ in the deep ultraviolet (UV), a theory of this type is weakly
coupled and can be described by perturbative methods.  As $\mu$ decreases, the
gauge coupling increases, as described by the renormalization group (RG) and
associated beta function.  To understand the infrared (IR) properties of a
strongly coupled chiral gauge theory has long been, and continues to be, an
outstanding goal in quantum field theory. If the theory satisfies the 't Hooft
global anomaly-matching conditions, then it might confine and produce massless
gauge-singlet composite spin-1/2 fermions \cite{thooft1979}-\cite{cgt}.
Alternatively, the strong gauge interaction could produce bilinear fermion
condensates. A chiral gauge theory that does not contain any vectorlike fermion
subsector is defined as being irreducibly chiral.  If a chiral gauge theory has
an irreducibly chiral fermion content, then these fermion condensates
necessarily break the chiral gauge symmetry
\cite{dfcgt,cgt},\cite{g78}-\cite{uvcomplete_etc}, whereas if it contains a
vectorlike fermion subsector, then condensates of fermions in this vectorlike
subsector may preserve the gauge symmetry. In both cases, the fermion
condensates break global chiral flavor symmetries.  In general, there can be
several stages of condensate formation at different momentum scales, with a
resultant sequence of gauge and/or global symmetry breaking. Here and below, we
restrict our consideration to asymptotically free chiral gauge theories that
have no anomalies in gauged currents, as is required for
renormalizability. Further, we restrict to theories with only gauge and fermion
fields but without any scalar fields.

There are several methods that one can use to investigate the ultraviolet to
infrared evolution of a chiral gauge theory. These include (i) (perturbative)
calculation of the beta function and analysis of possible IR zeros of this beta
function; (ii) use of the most-attractive-channel (MAC) approach, which can
suggest in which channel(s) bilinear fermion condensates are most likely to
form \cite{mac} if the coupling gets sufficiently strong in the infrared; and
(iii) a conjectured inequality involving the perturbative degrees of freedom in
the massless fields \cite{dfvgt,dfcgt}. We will denote this as the conjectured
DFI, where DFI stands for \underline{d}egree of \underline{f}reedom
\underline{i}nequality.  As was shown in \cite{dfcgt} and discussed further in
\cite{ads,cgt}, if the types of UV to IR evolution involving either formation
of fermion condensates with associated spontaneous chiral gauge and global
symmetry breaking or confinement with production of massless composite fermions
were to occur over a sufficiently large range of fermion contents
(specifically, a sufficiently large range of values of $p$ in the $Sp$ model
reviewed in Sect. \ref{spmodel}), these would violate the conjectured
degree-of-freedom inequality.  Hence, assuming the validity of the conjectured
degree-of-freedom inequality imposes significant restrictions on the behaviors
of these theories.  Moreover, as noted in \cite{cgt}, the type of UV to IR
evolution that would obey the degree-of-freedom inequality over the greatest
range of $p$ values is not the one favored by the MAC approach.  These results
lead one to inquire whether it is possible to achieve the goal of constructing
chiral gauge theories where the expected type(s) of UV to IR evolution obey the
conjectured degree-of-freedom inequality throughout the full range of
parameters specifying the fermion contents of these theories.

In this paper we report a successful achievement of this goal and give several
examples of such theories. Our theories have the gauge group ${\rm SU}(N)$ and
massless chiral fermions transforming according to a symmetric rank-2 tensor
representation of SU($N$), denoted $S$, and $N+4$ copies (i.e., flavors) of a
conjugate fundamental representation, denoted $\bar F$, together with a
vectorlike subsector consisting of $p$ copies of massless chiral fermions in
higher-dimensional representation(s).  Because SU(2) has only (pseudo)real
representations, it does not yield a chiral gauge theory, so we restrict our
considerations to chiral gauge theories having a gauge group SU($N$) with $N
\ge 3$.  We construct and analyze three theories of this type.  In the first
two, the higher-dimensional representation $R$ of the fermions in the
vectorlike subsector is self-conjugate, i.e., $R = \bar R$. These theories have
$p$ copies of chiral fermions in (a) the adjoint representation, $Adj$, and (b)
for $N=2k$ even, $p$ copies of chiral fermions in the $k$-fold antisymmetric
tensor representation, denoted $[N/2]_N=[k]_{2k}$.  For properties that are
common to both of these two theories, we will use the generic symbol $R_{sc}$
to refer to the respective self-conjugate ($sc$) representations. In the third
type of theory, (c), the vectorlike subsector is comprised of $p$ copies of
pairs of fermions of the form $\{R + \bar R \}$ with $R=S$.  Each of these
three types of chiral gauge theories thus consists of an irreducibly chiral
subsector, namely the $S$ and $N+4$ copies of $\bar F$ fermions, together with
a vectorlike subsector. Although we shall refer to these as three theories,
each one is really a two-parameter class of theories depending on $N$ and $p$.

We have chosen the representation $R$ of the fermions in the vectorlike
subsector of the theories studied in this paper so that for values of $N$ and
$p$ that lead to sufficiently strong gauge coupling in the infrared and
associated formation of bilinear fermion condensates, the most attractive
channel for condensation involves the fermions in the vectorlike subsector and
is of the form $R \times \bar R \to 1$, where here, the symbol 1 denotes a
singlet under SU($N$). This contrasts with the theory studied in
\cite{dfcgt,ads,cgt}, which has a vectorlike subsector consisting of $p$ copies
of massless fermions transforming as $\{F + \bar F\}$.  In that theory, the
most attractive channel is $S \times \bar F \to F$ rather than $F \times \bar F
\to 1$.  For each of our new chiral gauge theories, we present results on beta
functions, IR zeros of the respective beta functions, and predictions from the
most attractive channel approach. We then demonstrate that in each theory, for
each type of expected UV to IR evolution, the conjectured degree-of-freedom
inequality is obeyed throughout the full parameter range.

If the gauge theory is irreducibly chiral, then the gauge invariance forbids
any fermion masses in the Lagrangian.  For our purposes we will assume that the
masses of the fermions in vectorlike subsector are also zero.  This assumption
does not entail a significant loss of generality, because if a fermion in the
vectorlike subsector had a nonzero mass $m$, then as the reference scale $\mu$
decreases below $m$, one would integrate this vectorlike fermion out of the
low-energy effective theory applicable below that scale, and the result for the
infrared behavior would be equivalent to a theory without this fermion.

This paper is organized as follows.  In Sect. \ref{methods} we discuss our
general theoretical framework and methods of analysis.  Sect. \ref{spmodel} is
devoted to a brief review of a theory studied previously in
\cite{dfcgt,ads,cgt}. In Sect. \ref{strategy} we explain the basic strategy
that we use to construct our chiral gauge theories.  In Sects. \ref{adjmodel}
and \ref{atmodel} we present and analyze two new chiral gauge theories with
vectorlike subsectors having fermions transforming according to self-conjugate
representations of the gauge group. In Sect. \ref{global_flavor_symmetry} we
discuss the global flavor symmetry group for these two types of theories.  For
the values of $N$ and $p$ that lead to strong coupling in the infrared and
fermion condensation, we then analyze, in Sect. \ref{leeft}, the further
evolution into the infrared of the low-energy effective field theory that is
applicable below the scale of this initial condensation.  In
Sect. \ref{dfi_comparison} we demonstrate that for both of these new chiral
gauge theories with a given SU($N$) gauge group, the expected UV to IR
evolution obeys the conjectured degree-of-freedom inequality for the full range
of values of $p$.  Section \ref{ssbarmodel} is devoted to the analysis of the
third type of chiral gauge theory, with the type-(c) vectorlike subsector.
Again, we show that the conjectured degree-of-freedom inequality is obeyed for
this theory.  Our conclusions are given in Sect. \ref{conclusions}, and
some relevant formulas are included in Appendix \ref{bn}.


\section{Theoretical Framework and Methods of Analysis}
\label{methods}

In this section we discuss the theoretical framework and methods of analysis
that we use. As noted above, we consider asymptotically free chiral gauge
theories with gauge group $G={\rm SU}(N)$ and denote the gauge coupling
measured at a Euclidean momentum scale as $g(\mu)$.  It is also convenient to
use the quantities $\alpha(\mu) = g(\mu)^2/(4\pi)$ and
\beq
a(\mu) \equiv \frac{g(\mu)^2}{16\pi^2} = \frac{\alpha(\mu)}{4\pi} \ . 
\label{a}
\eeq 
(The argument $\mu$ in these couplings will often be suppressed in the
notation.)  Without loss of generality, we write all fermion fields in terms of
left-handed chiral components.  


\subsection{Beta Function} 
\label{betafunction}

The ultraviolet to infrared evolution of the gauge coupling is described by the
beta function, $\beta_g = dg/dt$, or equivalently,
\beq
\beta_\alpha = \frac{d\alpha}{dt} = \frac{g}{2\pi} \, \beta_g
\label{betaa}
\eeq
where $dt = d\ln \mu$. This has the series expansion
\beq
\beta_\alpha = -2\alpha \sum_{\ell=1}^\infty b_\ell \, a^\ell =
-2\alpha \sum_{\ell=1}^\infty \bar b_\ell \, \alpha^\ell \ ,
\label{beta}
\eeq
where we have extracted an overall minus sign, $b_\ell$ is the $\ell$-loop
coefficient, and $\bar b_\ell = b_\ell/(4\pi)^\ell$.  The $n$-loop beta
function, denoted $\beta_{\alpha,n\ell}$, is given by Eq. (\ref{beta}) with the
upper limit on the $\ell$-loop summation equal to $n$ instead of $\infty$.  The
property of asymptotic freedom means that $\beta_\alpha < 0$ for small
$\alpha$.  With the minus sign extracted in the perturbative expansion
(\ref{beta}), this is satisfied if $b_1 > 0$.  The one-loop and two-loop
coefficients $b_1$ \cite{b1} and $b_2$ \cite{b2} are independent of the scheme
used for regularization and renormalization, while the $b_\ell$ with $\ell \ge
3$ are scheme-dependent.

If $b_2 < 0$, then the two-loop beta function, $\beta_{\alpha,2\ell}$, has an
IR zero at
\beq
\alpha_{_{IR,2\ell}} = 4\pi a_{_{IR,2\ell}}=-\frac{4\pi b_1}{b_2} \ . 
\label{alfir_2loop}
\eeq
For sufficiently small fermion content, $b_2$ is positive, but as one enlarges
the fermion content in the theory, the sign of $b_2$ can become negative while
the theory is still asymptotically free, yielding an infrared zero in
$\beta_{\alpha,2\ell}$ at the above value.  If a theory has such an infrared
zero in the beta function, then, as the reference scale $\mu$ decreases from
large values in the ultraviolet, $\alpha(\mu)$ increases toward this infrared
zero.  If this IR zero occurs at sufficiently weak coupling, one expects that
the theory evolves from the UV to the IR without confinement or spontaneous
chiral symmetry breaking (S$\chi$SB), to a non-Abelian Coulomb phase.  In this
case, the infrared zero of beta is an exact IR fixed point (IRFP) of the
renormalization group, and as $\mu \to 0$ and the beta function vanishes, and
the theory exhibits scaling behavior with nonzero anomalous dimensions.  This
phenomena was discussed for vectorial gauge theories in \cite{b2,bz}.


\subsection{Most-Attractive-Channel Approach} 
\label{macsection}

In a theory whose UV to IR evolution leads to a gauge coupling that is strong
enough to produce fermion condensates, one method that has been widely used to
predict which type of condensate is most likely to form is the
most-attractive-channel (MAC) approach \cite{mac}.  Let us consider a
condensation channel in which fermions in the representations $R_1$ and $R_2$
of a given gauge group form a condensate that transforms according to the
representation $R_{cond.}$ of this group, denoted
\beq
\quad R_1 \times R_2 \to R_{cond.} \ . 
\label{channel}
\eeq
An approximate measure, based on
one-gluon exchange, of the attractiveness of this condensation channel, is 
\beq
\Delta C_2 = C_2(R_1) + C_2(R_2) -C_2(R_{Ch}) \ ,
\label{deltac2}
\eeq
where $C_2(R)$ is the quadratic Casimir invariant for the representation $R$
\cite{casimir}. At this level of one-gluon exchange, if $\Delta C_2$ is
positive (negative), then the channel is attractive (repulsive). The most
attractive channel is the one that yields the maximum (positive) value of
$\Delta C_2$. The MAC approach predicts that if, {\it a priori}, several
condensation channels could occur, then the one that actually occurs is the
channel that has the largest (positive) value of $\Delta C_2$.  The MAC method
was applied, for example, in efforts to build reasonably UV-complete models
with dynamical electroweak symmetry breaking \cite{uvcomplete_etc}. These
models made use of asymptotically free chiral gauge interactions that became
strongly coupled, naturally leading to the formation of certain condensates (of
fermions subject to the chiral gauge interaction) in a hierarchy of scales
corresponding, via inverse powers, to the observed generational hierarchy of
Standard-Model fermion mass scales. In these previous applications of the MAC
approach, and also in our present application, one bears in mind that the MAC
method is based on the one-gluon exchange and hence is only a rough guide to
the nonperturbative phenomenon of fermion condensation.

An analysis of the Schwinger-Dyson equation for the propagator of a massless
fermion transforming according to the representation $R$ of a gauge group $G$
shows that, in the ladder (i.e., iterated one-gluon exchange) approximation the
minimum value of $\alpha$ for which fermion condensation occurs in a vectorial
gauge theory is given by the condition that $3\alpha_{cr} C_2(R)/\pi = 1$, or
equivalently, $3\alpha_{cr} \Delta C_2 /(2\pi) = 1$, since $\Delta C_2=2C_2(R)$
in this case \cite{chipt}.  Therefore, an estimate is that as $\mu$
decreases and $\alpha(\mu)$ increases, condensation will first occur in a given
channel $Ch$ when $\alpha(\mu)$ increases through a critical value
\beq
\alpha_{cr,Ch} \sim \frac{2\pi}{3 \Delta C_2(R)_{Ch}} \ ,
\label{alfcrit}
\eeq
where we have labelled $C_2(R)$ with a subscript for the channel $Ch$.  This
estimate will be of particular interest for the most attractive channel.
Clearly, because of the strong-coupling nature of the fermion condensation
process, Eq. (\ref{alfcrit}) is only a rough estimate.  A measure of the
likelihood that the coupling grows large enough in the infrared to produce
fermion condensation in a given channel $Ch$ is the ratio
\beq
\rho_{_{IR,Ch}} \equiv \frac{\alpha_{_{IR,2\ell}}}{\alpha_{cr,Ch}} \ .
\label{ratio}
\eeq
If this ratio is significantly larger (smaller) than unity, one may infer that
condensation in the channel $Ch$ is likely (unlikely).  As with the caveats
given above concerning the MAC, in using this ratio $\rho_{_{IR,Ch}}$, one is 
cognizant of the theoretical uncertainties due to the strong-coupling nature of
the physics.


\subsection{Degree-of-Freedom Inequality} 
\label{dfi_section}

A quantity that can give interesting predictions for renormalization-group
evolution involves the relevant perturbative field degrees of freedom in the
effective field theory that is applicable at a given reference scale, $\mu$.
From the study of second-order phase transitions and critical phenomena in
statistical mechanics and condensed matter physics, one is familiar with the
Wilsonian thinning of degrees of freedom as one changes the scale on which one
measures physical quantities from short distances (UV) to large distances (IR).
Given the correspondence between the inverse distance and the reference
momentum scale $\mu$, one may naturally expect a similar decrease (or
non-increase) of dynamical degrees of freedom in a quantum field theory as
$\mu$ decreases from large values in the ultraviolet to small values in the
infrared. In conformal field theory in $d=2$ dimensions, it has been proved
that a certain quantity that can be interpreted as a measure of the degrees of
freedom (the central charge of the associated Virasoro algebra) decreases as a
function of the renormalization-group flow \cite{zam}.

Given that a theory is asymptotically free, the gauge coupling approaches zero
in the deep ultraviolet as $\mu \to \infty$, so that one can identify and
enumerate the perturbative degrees of freedom in the fields.  Depending on the
theory, it may also be true that in the deep infrared, as $\mu \to 0$, the
residual (massless) particles are weakly interacting, so that again one can
describe them perturbatively and enumerate their degrees of freedom.  Although
one is describing the UV to IR evolution of a zero-temperature quantum field
theory, a natural approach to the enumeration of the perturbative degrees of
freedom in the fields is provided by envisioning a finite-temperature field
theory, where the temperature $T$ corresponds to the Euclidean scale, $\mu$,
and using the count embodied in the free energy density, $F(T)$.  This is given
by
\beq
F(T) = f(T) \, \frac{\pi^2}{90} \, T^4
\label{free_energy}
\eeq
with 
\beq
f = 2N_v + \frac{7}{4}N_f + \frac{7}{8}N_{f,Maj} + N_s \ ,
\label{f}
\eeq
where $N_v$ and $N_s$ are the number of vector and (real) scalar fields, and
$N_f$ and $N_{f,Maj}$ are the number of chiral components of Dirac and Majorana
fermions in the theory, respectively \cite{acon,sc}. Assuming that the relevant
fields become free in the respective UV and IR limits, we define
\beq
f_{UV} = f(\infty), \quad f_{IR} = f(0) \ . 
\label{fuvir}
\eeq
Since the theories that we consider are required to be asymptotically free, we
can always identify the Lagrangian fields in the deep UV and hence calculate
$f_{UV}$.  

In accord with experience in statistical mechanics, Ref. \cite{dfvgt}
conjectured the degree-of-freedom inequality
\beq
\Delta f \equiv f_{UV} - f_{IR} \ge 0 
\label{dfi}
\eeq
for vectorial gauge theories, and Ref. \cite{dfcgt} extended this conjecture to
chiral gauge theories.  In \cite{dfcgt} this conjecture was applied to analyze
several asymptotically free chiral gauge theories. Subsequent studies have
investigated the possible types of IR behavior involving strong coupling and
condensate formation; Refs. \cite{ads,cgt} are particularly relevant for
our current work.  

As noted above, since we restrict to asymptotically free theories, the
condition that the theory becomes free as $\mu \to \infty$ is always satisfied.
There are three types of situations where the condition that the fields are
also weakly coupled in the IR is satisfied.  In all of these we can calculate
$f_{IR}$.  In the first of these, the theory evolves to an exact, weakly
coupled IR fixed point, so that the field degrees of freedom in the massless
fields are the same as they were in the UV, up to small, calculable
perturbative corrections, which obey the inequality (\ref{dfi})
\cite{dfvgt,dfcgt}. In the second type of situation, there is global and/or
gauge symmetry breaking at one or more scales, so that as $\mu$ decreases below
these scales toward the infrared, in the applicable low-energy effective field
theory, the remaining massless particles are Nambu-Goldstone bosons (NGBs)
resulting from the spontaneous chiral symmetry breaking. Since the NGBs have
only derivative interactions among themselves, which vanish as
$\sqrt{s}/\Lambda \to 0$, where $\sqrt{s}$ is the center-of-mass energy and
$\Lambda$ denotes the scale of chiral symmetry breaking, it follows that these
NGBs become free in the infrared limit. A third type of possible situation is
one in which the chiral gauge interaction confines and produces massless
gauge-singlet composite fermions.  The interactions between these gauge-singlet
fermions involve higher-dimension operators and hence are also weak in the
infrared.  In some models, the second and third types of behavior can occur
together \cite{cgt}.

A direct test of the conjectured degree-of-freedom inequality (\ref{dfi}) for
asymptotically free chiral gauge theories would probably require lattice
simulations.  However, because of fermion doubling on the lattice (in which a
single continuum fermion produces $2^d$ fermion modes on a $d$-dimensional
Euclidean lattice, with half corresponding to one sign of $\gamma_5$ and the
other half corresponding to the opposite sign of $\gamma_5$), it has been
challenging to simulate chiral gauge theories via lattice methods.  A different
approach to testing the validity of the conjecture is to study its application
to vectorial gauge theories.  These have the advantage that they can be
simulated on the lattice, and there are well-understood ways of dealing with
fermion doubling so that in the continuum limit one should be able to determine
the actual number, $N_f$, of active fermions.  Ongoing lattice studies of the
infrared behavior of various vectorial gauge theories, such as a gauge theory
with $G={\rm SU}(2)$ and $N_f=6$ Dirac fermions in the fundamental
representation \cite{su2nf6}, are making progress in testing the conjectured
degree-of-freedom inequality.


\section{The $Sp$ Theory}
\label{spmodel}

In this section we review the properties of a chiral gauge theory that has been
studied before \cite{by,by2,dfcgt,ads,cgt} and provides motivation for our
present work.  The reader who is familiar with this material could skip this
section and proceed to Sect. \ref{strategy}.  This theory, which we denote the
$Sp$ model, has the gauge group SU($N$) and massless chiral fermions
transforming according to
\begin{enumerate}

\item 
a symmetric rank-2 tensor representation, $S$, with corresponding field
$\psi^{ab}_L = \psi^{ba}_L$,

\item 
$N+4$ copies of chiral fermions in the conjugate fundamental representation,
$\bar F$, with fields $\chi_{a,i,L}$, $i=1,...,N+4$, and

\item 
a vectorlike subsector consisting of $p$ copies of pairs of chiral fermions
transforming as $\{F + \bar F\}$, with fields $\chi^a_{j,L}$ and
$\chi_{a,j,L}$, $j=1,...,p$.  

\end{enumerate}
Here and below, $a,b,c...$ are gauge indices and $i,j,..$ are copy (i.e., 
flavor) indices. 

The one- and two-loop coefficients in the beta function of this theory are 
\beq
(b_1)_{Sp} = 3N-2-\frac{2p}{3} 
\label{b1_spmodel}
\eeq
and
\beq
(b_2)_{Sp}=\frac{13}{2}N^2 - 15N + \frac{1}{2} + 6N^{-1} + 
p\Big ( -\frac{13N}{3} + N^{-1} \Big ) \ . 
\label{b2_spmodel}
\eeq
The coefficient $(b_1)_{Sp}$ decreases with $p$ and vanishes at 
$p = p_{b1z,Sp} = (9/2)N-3$, 
where the subscript $bnz$ stands for ``$b_n$ equals \underline{z}ero''.
Asymptotic freedom requires $(b_1)_{Sp} > 0$, i.e., 
\beq
p < \frac{9}{2}N-3 \ . 
\label{p_upperbound_spmodel}
\eeq
The two-loop coefficient is positive for small $p$ and decreases through zero
to negative values as $p$ increases through the value 
\beq
p_{b2z,Sp} = \frac{3(13N^3-30N^2+N+12)}{2(13N^2-3)} \ .
\label{pb2z_spmodel}
\eeq
In the interval
\beq
(I_p)_{Sp}: \quad p_{b2z,Sp} < p < p_{b1z,Sp}
\label{pinterval_spmodel}
\eeq
the two-loop beta function has an infrared zero, which occurs at the value
\beqs
& & \alpha_{_{IR,2\ell,Sp}} = 
\frac{8\pi N(9N-6-2p)}{p(26N^2-6)-39N^3+90N^2-3N -36} \ .  \cr\cr
& & 
\label{alfir_2loop_spmodel}
\eeqs
Clearly, the two-loop perturbative calculation that yields this result
(\ref{alfir_2loop_spmodel}) is most accurate if $p$ is near the upper end of
the interval $(I_p)_{Sp}$, where $\alpha_{_{IR,2\ell,Sp}}$ is small, and
becomes less reliable as $p$ approaches the lower end of the interval
$(I_p)_{Sp}$. 

For this theory, the most attractive channel for fermion condensation is 
\beq
S \times \bar F \to F \ , 
\label{sfbarchannel}
\eeq
with attractiveness measure
\beq
\Delta C_2 = C_2(S) = \frac{(N+2)(N-1)}{N} \quad {\rm for} \ 
S \times \bar F \to F \ . 
\label{deltac2_sfbar}
\eeq
Hence, for this channel, 
\beqs
\rho_{_{IR,S \times \bar F}} & \equiv & 
\frac{\alpha_{_{IR,2\ell,Sp}}}{\alpha_{cr,S \times \bar F}} 
\cr\cr 
& = & \frac{12(9N-6-2p)(N+2)(N-1)}{p(26N^2-6)-39N^3+90N^2-3N-36} \ . 
\cr\cr
& & 
\label{r_spmodel}
\eeqs
This ratio exceeds unity for 
\beq
p <  p_{cr,Sp} = \frac{3(49N^3-18N^2-95N+60)}{2(25N^2+12N-27)} \ . 
\label{pcondense_spmodel}
\eeq

If $p$ is only slightly less than $p_{b1z}$, then $\rho_{_{IR,S \times \bar F}}
\ll 1$, so the UV to IR evolution is expected to be to a deconfined, weakly
coupled non-Abelian Coulomb phase. Here, also taking into account perturbative
corrections to the free-field count of field degrees of freedom, the DFI is
obeyed \cite{dfcgt}.

If $p$ is sufficiently small (with either $ p \in (I_p)_{Sp}$ or $1 \le p <
p_{b2z,Sp}$), then the theory becomes strongly coupled in the infrared.  For
these values of $p$, one possible type of UV to IR evolution could produce
confinement with massless, gauge-singlet composite fermions \cite{by} and no
spontaneous chiral symmetry breaking. Alternately, there could be fermion
condensation in the most attractive channel (\ref{sfbarchannel}), breaking the
gauge group SU($N$) to SU($N-1$) and also breaking global flavor symmetries.
The associated fermion condensate has the form
\beq
\langle \psi^{ab \ T}_L C \chi_{b,i,L}\rangle  \ . 
\label{condensate_spmodel}
\eeq
Without loss of generality, one may pick $a=N$ and $i=1$.  The fermions
involved in this condensate gain dynamical masses, and one then constructs the
low-energy effective field theory applicable at lower scales.  The coupling in
this low-energy theory continues to grow and is again expected to produce a
condensate in the most attractive channel, $S \times \bar F$, where now $S$ and
$\bar F$ refer to representations of SU($N-1$).  This process continues
sequentially until the original SU($N$) gauge symmetry in the UV is completely
broken.

The degree-of-freedom measure in the UV is 
\beqs
& & f_{UV,Sp} = 2(N^2-1) + \frac{7}{4}
\Big [ \frac{N(N+1)}{2} + (N+4+2p)N \Big ] \ . \cr\cr
& & 
\label{fuv_spmodel}
\eeqs
For the possible type of UV to IR evolution that leads to confinement and
massless composite fermions (labelled with the subscript $sym$), one finds
\cite{dfcgt}
\beqs
f_{IR,Sp;sym} & = & \frac{7}{4}\Big [ \frac{1}{2}(N+4+p)(N+3+p) \cr\cr
& + & p(N+4+p) + \frac{1}{2}p(p+1) \Big ] \ . \cr\cr
& & 
\label{fir_spmodel_sym}
\eeqs
Here and below, the subscripts after $IR$ in a quantity such as $f_{IR,Sp;sym}$
refer to the theory (here, the $Sp$ theory) and then, after the semicolon, the
type of UV to IR evolution.  Thus, for this type of UV to IR flow,
\beqs
(\Delta f)_{Sp;sym} & \equiv & f_{UV,Sp} - f_{IR,Sp;sym} \cr\cr
& = & \frac{1}{4}\Big [ 15N^2+7N-50-14p(4+p) \Big ] \ . \cr\cr
& & 
\label{deltaf_spmodel_sym}
\eeqs
(Here, in the symbol $(\Delta f)_{Sp;sym}$, the first subscript identifies the
theory and the subscripts after the semicolon identify the type of UV to IR
evolution; the same notation is used for the other theories to be discussed.)
The difference $(\Delta f)_{Sp;sym}$ is positive if and only if
\beq
p < -2 + \sqrt{ \frac{15N^2+7N+6}{14}} \ . 
\label{p_upperbound_dfi_sym_spmodel}
\eeq

For the type of UV to IR flow involving sequential fermion 
condensation in the $S \times \bar F \to F$ channels
\beqs
f_{IR,Sp;S \times \bar F} & = & 2N(4+p)+1 \cr\cr
& + & \frac{7}{4}\Big [ \frac{N(N-1)}{2} + 4N + 2pN \Big ] \ . \cr\cr
& & 
\label{fir_spmodel_sfbar}
\eeqs
Consequently, for this type of UV to IR flow, 
\beqs
(\Delta f)_{Sp;S \times \bar F} & \equiv & f_{UV,Sp}-f_{IR,Sp;S \times \bar F}
 \cr\cr
& = & \frac{1}{4}\Big [ 15N^2-25N-12-8pN \Big ] \ . \cr\cr
& & 
\label{deltaf_spmodel_sfbar}
\eeqs
This is positive if and only if
\beq
p < \frac{15N^2-25N-12}{8N} \ . 
\label{p_upperbound_dfi_bk_spmodel}
\eeq

If, for a given $N$, the upper bounds (\ref{p_upperbound_dfi_sym_spmodel}) and
(\ref{p_upperbound_dfi_bk_spmodel}) were substantially greater than the value
of $p_{cr,Sp}$ in Eq. (\ref{pcondense_spmodel}), then they would not be
important, since in this region, toward the upper end of the interval
$(I_p)_{Sp}$, one would expect that the UV to IR evolution would be to a
deconfined non-Abelian Couolmb phase, for which the conjectured DFI is obeyed.
However, these upper bounds (\ref{p_upperbound_dfi_sym_spmodel}) and
(\ref{p_upperbound_dfi_bk_spmodel}) are less then $p_{cr,Sp}$.  For example,
for $N=3$, we have $p_{b2z,Sp}=24/19=1.263$ (to the given floating point
accuracy) and $p_{b1z,Sp}=21/2=10.5$, so the interval $(I_p)_{Sp}$ consists of
the values $2 \le p \le 10$. Furthermore, for this $N=3$ value,
$p_{cr,Sp}=6$, so that for $p \lsim 6$, one may anticipate
that the UV to IR evolution would plausibly involve strong coupling, as
embodied in the two types of evolution discussed above, namely confinement with
massless composite fermions and no spontaneous chiral symmetry breaking or
production of fermion condensates and associated gauge and global symmetry
breaking. Now
\beq
N=3 \ \Rightarrow (\Delta f)_{Sp;sym} > 0 \quad {\rm if} \ \ 
p < \frac{-14+9\sqrt{7}}{7} =1.402 \ , 
\label{Deltaf_Sp_sym_Nc3}
\eeq
so that if this UV to IR evolution leading to massless composite fermions
without any spontaneous chiral symmetry breaking were to occur for values in
the strongly coupled range of $p$, $2 \le p \lsim 6$, then it would violate the
conjectured degree-of-freedom inequality (\ref{dfi}).  Furthermore,
\beq
N=3 \ \Rightarrow (\Delta f)_{Sp;S \times \bar F} > 0 \quad {\rm if} \ 
p < 2 \ .
\label{Deltaf_Sp_bk_Nc3}
\eeq
Hence, if the UV to IR evolution were to lead to condensate formation in the
successive $S \times \bar F$ channels of the SU($N$) theory, the SU($N-1$)
theory, etc., then it would violate the conjectured DFI for much of the
strongly-coupled range of values of $p$, including $2 \le p \lsim 6$.

In general, the $Sp$ model is a two-parameter theory, depending on both $N$ and
$p$. An interesting limit is 
\beqs
& & LNP: \quad N \to \infty \ , \quad p \to \infty \cr\cr
& & {\rm with} \ r \equiv \frac{p}{N} \ {\rm fixed} \quad
{\rm and} \quad \alpha(\mu) N \ {\rm finite} \ . \cr\cr
& &
\label{lnp}
\eeqs
We denote this as the LNP (large $N$ and $p$) limit \cite{lnlnp}. In this LNP
limit, the resultant theory evidently depends only on the single parameter
$r$. We define
\beq
r_{bnz} \equiv \lim_{LNP} \frac{p_{bnz}}{N} \ , n=1,2
\label{rbnz}
\eeq
One has 
\beq
r_{b1z} = \frac{9}{2}
\label{rb1z_lnp}
\eeq
and
\beq
r_{b2z} = \frac{3}{2} 
\label{rb2z_lnp}
\eeq
so that the analogue of $(I_p)_{Sp}$ for this LNP limit is 
\beq
(I_r)_{Sp}: \quad 1.5 < r < 4.5 \ .
\label{rinterval_sp_lnp}
\eeq
Further, 
\beq
r_{cr,S \times \bar F} \equiv \lim_{LNP} \frac{p_{cr,S \times \bar F}}{N}
                  = \frac{147}{50} = 2.94  \ . 
\label{rcrit_sfbar_lnp}
\eeq

We define a rescaled degree-of-freedom measure that is finite in the LNP limit,
namely 
\beq
\bar f \equiv \lim_{LNP} \frac{f}{N^2} \ . 
\label{fbar_lnp}
\eeq
One has 
\beq
\bar f_{UV,Sp} =\frac{37}{8} + \frac{7}{2}r 
\label{fuvbar_spmodel}
\eeq
\beq
\bar f_{IR,Sp;sym} = \frac{7}{8}+\frac{7}{2}r(1+r) 
\label{firbar_spmodel_sym_lnp}
\eeq
and
\beq
\bar f_{IR,Sp; S \times \bar F} = \frac{7}{8} + \frac{11}{2}r \ . 
\label{firbar_spmodel_sfbar_lnp}
\eeq
Consequently, for the type of UV to IR evolution that leads to confinement and
massless composite fermions, which might occur in the strongly coupled IR
regime where $r \lsim 3$, 
\beq
(\Delta \bar f)_{Sp;sym} \equiv \bar f_{UV,Sp} - \bar f_{IR,Sp;sym} = 
\frac{15-14r^2}{4}  \ . 
\label{deltaf_spmodel_sym_lnp}
\eeq
This would obey the conjectured DFI only if \cite{dfcgt,cgt,cgtmp} 
\beq
r < \sqrt{\frac{15}{14}} = 1.035 \ . 
\label{r_upperbound_dfi_sym_spmodel_lnp}
\eeq
For the possible type of UV to IR evolution that leads to sequential fermion
condensation in the $S \times \bar F \to F$ channels,
\beq (\Delta \bar f)_{Sp;S \times \bar F} \equiv f_{UV,Sp}-\bar f_{IR,Sp;S
\times \bar F} = \frac{15-8r}{4} \ .
\label{deltaf_spmodel_sfbar_lnp}
\eeq
This would obey the conjectured DFI only if
\beq
r < \frac{15}{8} = 1.875 \ . 
\label{r_upperbound_dfi_sfbar_spmodel_lnp}
\eeq
Both of the upper limits (\ref{r_upperbound_dfi_sym_spmodel_lnp}) and
(\ref{r_upperbound_dfi_sfbar_spmodel_lnp}) are well below the upper bound from
asymptotic freedom, $r < 4.5$.  Importantly, they are also below the value of
$r \sim 3$ where the estimate Eq. (\ref{rcrit_sfbar_lnp}) suggests that
strong-coupling behavior occurs.  Hence, in this $Sp$ model, there is
considerable uncertainty in the overall prediction for the UV to IR evolution
in the case where this involves strong coupling.  Assuming the validity of the
conjectured degree-of-freedom inequality, this DFI would forbid two types of
strongly coupled UV to IR evolution that would otherwise be inferred to be
likely, namely confinement without any spontaneous chiral symmetry breaking in
the interval $\sqrt{15/14} < r \lsim 3$ and condensate formation in the MAC
with attendant gauge and chiral symmetry breaking in the interval $15/8 < r
\lsim 3$.

This property of the $Sp$ model, noted in \cite{dfcgt} and further discussed in
\cite{ads,cgt} provides a motivation for the goal of constructing
asymptotically free chiral gauge theories where the likely type(s) of UV
to IR evolution is (are) in agreement with the conjectured degree-of-freedom
inequality for the full range of fermion contents (as parametrized here by the
value of $p$). We have achieved this goal, as we report in the present work.

We include a parenthetical remark here.  In the $Sp$ model, although not
favored by the MAC criterion, if there were condensation of the fermions in the
vectorlike subsector in the $F \times \bar F \to 1$ channel, followed at lower
scales by either confinement with massless composite fermions or sequential
condensate formation in the $S \times \bar F \to F$ channels, then
\beqs
\bar f_{IR,Sp;F \times \bar F,sym} & = & 
\bar f_{IR,Sp;F \times \bar F, S \times \bar F} \cr\cr
& = & \frac{7}{8} + r(2+r) \ .
\label{firbar_spmodel_ffbar_sfbar}
\eeqs
Hence, for this type of UV to IR evolution, 
\beqs
(\Delta \bar f)_{Sp;F \times \bar F, S \times \bar F} & \equiv &
\bar f_{UV,Sp} - \bar f_{IR,Sp; F \times \bar F, S \times \bar F} \cr\cr
& = & \frac{1}{4}(15+6r-4r^2) \ . 
\label{deltafbar_spmodel_ffbar_sfbar}
\eeqs
This is positive for 
\beq
r < \frac{3+\sqrt{69}}{4} = 2.83 \ . 
\label{rffbarsfbar}
\eeq
Thus, of the various possible types of UV to IR evolution in the $Sp$ theory,
the type that obeys the DFI conjecture over the largest range of $r$, is
condensation in the $F \times \bar F \to 1$ channel, followed at lower scales
by either confinement with massless composite fermions or sequential condensate
formation in the $S \times \bar F \to F$ channels. But the initial condensation
in this type of evolution is not the one favored by the MAC criterion, which,
instead favors initial and then sequential condensation in the $S \times \bar F
\to F$ channels of the SU($N$) then $S \times \bar F \to F$ condensation in the
SU($N-1$) theory, and so forth, until the SU($N$) gauge symmetry is completely
broken.


\section{Strategy for Construction of New Chiral Gauge Theories}
\label{strategy}

Our general method for constructing the chiral gauge theories presented here is
as follows. We take the gauge group to be $G= {\rm SU}(N)$ and include,
as the irreducibly chiral sector of the theory, fermions transforming as the
$S$ and $(N+4)$ copies of $\bar F$.  We choose the vectorlike subsector to
consist of $p$ copies of fermions that transform according to representation(s)
$R$ of $G$ such that the channel
\beq
R \times \bar R \to 1
\label{rrbarto1_channel}
\eeq
is more attractive than other channels. (For some of our theories, $R = \bar
R$.) In the theories that we consider, the next-most-attractive channel is 
\beq
S \times \bar F \to F \ . 
\label{sfbartof_channel}
\eeq
The $\Delta C_2$ attractiveness measures for these channels are
\beq
\Delta C_2 = 2C_2(R) \quad {\rm for} \ R \times \bar R \to 1
\label{deltac2_rrbarto1_channel}
\eeq
and
\beq
\Delta C_2 = C_2(S) = \frac{(N+2)(N-1)}{N} \quad {\rm for} \ 
S \times \bar F \to \bar F \ , 
\label{deltac2_sfbar_channel}
\eeq
so the condition that the $ R \times \bar R \to 1$ channel is more attractive
than the $S \times \bar F \to \bar F$ channel is that 
\beq
\Delta C_2(R) = 2C_2(R) > \frac{(N+2)(N-1)}{N} \ . 
\label{deltac2_inequality}
\eeq
In all the cases that we consider, this guarantees that the $R \times \bar R
\to 1$ channel is the most attractive channel in which condensation thus occurs
first as the theory evolves from the UV to the IR.  Consequently, if the
fermion content is such that the running coupling $\alpha(\mu)$ becomes
sufficiently large in the infrared, then, because the MAC is
(\ref{rrbarto1_channel}), the fermion condensation at the highest energy scale
occurs among the fermions in the vectorlike subsector of the model, via the
channel $R \times \bar R \to 1$.  The resultant low-energy effective field
theory applicable below this scale is thus comprised of the irreducible chiral
sector of the theory, equivalent to the $p=0$ special case of the full theory,
with just the $S$ fermion and the $N+4$ copies of the $\bar F$ fermion.  As
reviewed in Sect. \ref{spmodel}, the various possible types of UV to IR
evolution of this $p=0$ theory obey the conjectured degree-of-freedom
inequality \cite{dfcgt,ads,cgt}.


\section{Theory with $R=Adj$} 
\label{adjmodel}


\subsection{Particle Content}
\label{adjmodel_content}

In this section we construct and study a chiral gauge
theory with gauge group SU($N$) and fermion content consisting of chiral
fermions transforming according to
\begin{enumerate}

\item 
a symmetric rank-2 tensor representation, $S$, with corresponding field
$\psi^{ab}_L = \psi^{ba}_L$,

\item 
$N+4$ copies (also called ``flavors'') of
chiral fermions in the conjugate fundamental representation, $\bar F$, with 
fields $\chi_{a,i,L}$, $i=1,...,N+4$, and 

\item 
$p$ copies of chiral fermions in the adjoint representation, denoted $Adj$,
  with fields $\xi^a_{b,j,L}$, $j=1,...,p$.

\end{enumerate}
Here and below, $a,b,c...$ are gauge indices and $i,j$ are copy indices.  We
call this the $Adj$ theory by reference to the choice of the representation
$R=R_{sc}$ for the fermions in the vectorlike subsector. This fermion content
is summarized in Table \ref{fermion_properties_rscmodels}. As noted above, we
restrict to $N \ge 3$ because SU(2) has only (pseudo)real representations and
hence a gauge theory based on the gauge group SU(2) is not chiral. This theory
thus depends on the two integer parameters, $N \ge 3$ and $p \ge 0$, with an
upper limit on $p$ given by Eq. (\ref{pb1z_adjmodel}) below. We will sometimes
use the Young tableaux $\sym$ and $\overline{\fund}$ for $S$ and $\bar F$.  The
irreducibly chiral sector of this theory is comprised of the $S$ and the $N+4$
copies of $\bar F$ fermions, and the vectorlike subsector is comprised of the
$Adj$ fermions. Because of this self-conjugate nature of $R_{sc}$, the $Adj$
fermions may be considered to be Majorana.  Thus, if one were to remove the
irreducibly chiral part of this theory and consider the part containing the
gauge fields and the $Adj$ fermions alone, the dynamical particle content in
the Lagrangian would be analagous to the gluons and gluinos of an ${\cal N}=1$
supersymmetric SU($N$) gauge theory.

We recall that since the contribution to the triangle anomaly from $S$
satisfies \cite{anom}
\beq
{\rm Anom}(S)=(N+4) \, {\rm Anom}(F) \ , 
\label{anom_sym2}
\eeq
and since 
\beq
{\rm Anom}(R)=-{\rm Anom}(\bar R) \ , 
\label{anomrrbar}
\eeq
it follows that the set of chiral fermions $S$ plus $(N+4)$ copies of $\bar F$
yields a theory that is free of anomalies in gauged currents. Furthermore, from
Eq. (\ref{anomrrbar}), it follows that for any self-conjugate representation
$R_{sc}$, $Anom(R_{sc}) = 0$.  Hence, we are free to add fermions transforming
according to a self-conjugate representation to a chiral gauge theory that is
free of anomalies in gauged currents and it will retain this anomaly-free
property. We use this fact here with $R_{sc}=Adj$.


\subsection{Beta Function}
\label{adjmodel_betafunction}

The beta function for this $Adj$ theory is given by Eq. (\ref{beta}) with the
one-loop coefficient
\beq
(b_1)_{Adj} = \frac{1}{3}\Big [ (9-2p)N -6 \Big ]
\label{b1_adjmodel}
\eeq
and the two-loop coefficient 
\beq
(b_2)_{Adj} = \frac{1}{6}\Big [ (39-32p)N^2-90N+3+36N^{-1} \Big ] \ . 
\label{b2_adjmodel}
\eeq
(See Appendix \ref{bn} for general formulas for $b_1$ and $b_2$.) These
coefficients contain the maximal scheme-independent information about the
dependence of the gauge coupling on the reference scale, $\mu$.  This
information will suffice for our present purposes.  Higher-loop effects for
vectorial theories and effects of scheme transformations on higher-loop terms
in the beta function for gauge theories have been studied in
\cite{hl}-\cite{srgt}.

We denote the values of $p$ for which $(b_1)_{Adj}=0$ as $p_{b1z,Adj}$ (where
the subscript stands for $b_1$ \underline{z}ero). This value is \cite{npreal}
\beq
p_{b1z,Adj} = \frac{3(3N-2)}{2N} \ . 
\label{pb1z_adjmodel}
\eeq
Our requirement that the model should be asymptotically free means that
$\beta_\alpha < 0$ for small $\alpha$. This is equivalent to the condition that
$b_1 > 0$ or, if $b_1$ vanishes, then the further requirement that $b_2 > 0$.
Now $(b_1)_{Adj} > 0$ if and only if $p < p_{b1z,Adj}$, i.e.,
\beq
p < \frac{3(3N-2)}{2N} \ . 
\label{p_upperbound_adjmodel}
\eeq
This means that the set of physical, integral values of $p$ allowed by our
requirement of asymptotic freedom are $0 \le p \le 3$ for $N=3, \ 4, \ 5, \ 6$
and $0 \le p \le 4$ for $N \ge 7$. Note that if $N=6$ and $p=4$, then $b_1=0$,
so one must examine the sign of $b_2$ to determine if the theory is
asymptotically free or not, and for this case $(b_2)_{Adj}$ is negative, hence
excluding it from consideration. Here and below, for a given theory and value
of $N$, we will denote the maximum allowed value of $p$ as $p_{max}$. 

As a consequence of the asymptotic freedom of the theory, the beta function
always has a zero at $\alpha=0$, which is a UV fixed point (UVFP) of the
renormalization group. In general, the two-loop beta function,
$\beta_{\alpha,2\ell}$, has an IR zero if $b_2$ has a sign opposite to that of
$b_1$, i.e., if $b_2$ is negative.  For $p=0$, $(b_2)_{Adj} > 0$, so
$\beta_{\alpha,2\ell}$ has no IR zero.  As $p$ increases, $(b_2)_{Adj}$
decreases and eventually passes through zero to negative values, giving rise to
an IR zero of $\beta_{\alpha,2\ell,Adj}$.  Let us denote the value of $p$ where
$b_2$ vanishes as $p_{b2z,Adj}$.  This is
\beq
p_{b2z,Adj} = \frac{3(13N^3-30N^2+N+12)}{32N^3} \ . 
\label{pb2z_adj}
\eeq
In Table \ref{pb12z_adjmodel} we list values of $p_{b1z,Adj}$ and $p_{b2z,Adj}$
for this theory. The value $p_{b2z,Adj}$ is less than the upper bound on 
$p$, $p_{b1z,Adj}$, i.e., 
\beq
p_{b2z,Adj} < p_{b1z,Adj} \ .
\label{pbinequality_adjmodel}
\eeq
This inequality is proved by analyzing the difference, 
\beq
p_{b1z,Adj} - p_{b2z,Adj} = \frac{3(35N^3-2N^2-N-12)}{32N^3} \ . 
\label{pbdif_adjmodel}
\eeq
This difference is positive for all physical $N$.  Hence, for 
$p$ in the interval \cite{npreal}
\beq
(I_p)_{Adj}: \quad p_{b2z,Adj} < p < p_{b1z,Adj} \ , 
\label{pinterval_adjmodel}
\eeq
this theory is asymptotically free, and $\beta_{\alpha,2\ell,Adj}$ has an IR
zero.  The actual physical, integral values of $p$ in the interval 
$(I_p)_{Adj}$ depend
on the value of $N$.  There are several different sets of $N$ and $p$ values
where this IR zero is physical:
\beqs
(I_p)_{Adj}: & \quad & 1 \le p \le 3 \quad {\rm if} \ 3 \le N \le 6,  \cr\cr
             &       & 1 \le p \le 4 \quad {\rm if} \ 7 \le N \le 12, \cr\cr
             &       & 2 \le p \le 4 \quad {\rm if} \ N \ge 13 \ . 
\label{pinterval_adjmodel_specific}
\eeqs
These different cases follow from two properties.  First, $p_{b1z,Adj}$
(continued to real numbers) is a monotonically increasing function of $N$ for
physical $N$ and ascends through the value 4 as $N$ increases through the value
$N=6$. Second, for $N > (1+\sqrt{1081} \ )/30 = 1.129$ and hence for the range
$N \ge 3$ relevant here, $p_{b2z,Adj}$ is a monotonically increasing function
and increases through 1 at $N=12.7922$ (the largest root of
$7N^3-90N^2+3N+36$).  Hence, if $N \ge 13$, the lowest value of $p \in
(I_p)_{Adj}$ is $p=2$, as indicated in (\ref{pinterval_adjmodel_specific}).

For values of $N$ and $p$ where $\beta_{\alpha,2\ell,Adj}$ has a physical IR
zero, it occurs at
\beqs
\alpha_{_{IR,2\ell,Adj}} & \equiv & 4\pi a_{_{IR,2\ell,Adj}} = 
-4\pi \frac{(b_1)_{Adj}}{(b_2)_{Adj}} \cr\cr
& & \cr
       & = & \frac{8\pi N[(9-2p)N-6]}{(32p-39)N^3+90N^2-3N-36} \ . 
\cr\cr
& & 
\label{air_2loop_adjmodel}
\eeqs
In using this result, it should be recalled that, in general, 
an IR zero of a beta function at $\alpha_{_{IR,2\ell}} = -4\pi b_1/b_2$ can be 
reliable if $|b_2|$ is not too small, i.e., when $\alpha_{_{IR,2\ell}}$ 
is not too large for the perturbative calculation to be applicable. 
In Table \ref{alfir_and_ratio_adjmodel} we list values of 
$\alpha_{_{IR,2\ell,Adj}}$.

It is of interest to consider the limit \cite{lnlnp} 
\beq
N \to \infty \ {\rm with} \ \zeta(\mu) \equiv \alpha(\mu) N \ 
{\rm finite \ and} \ p \ {\rm fixed}.
\label{ln}
\eeq
In this limit, 
\beq
\lim_{N \to \infty} p_{b1z,Adj} = \frac{9}{2} 
\label{pb1z_adjmodel_largeN}
\eeq
and 
\beq
\lim_{N \to \infty} p_{b2z,Adj} = \frac{39}{32} = 1.21875 \ , 
\label{pb2z_adjmodel_largeN}
\eeq
so that the interval $(I_p)_{Adj}$ becomes 
\beq
\lim_{N \to \infty} (I_p)_{Adj} : \ \frac{39}{32} < p < \frac{9}{2} \ , 
\label{pinterval_adjmodel_largeN}
\eeq
containing the physical, integral values $p=2, \ 3, \ 4$.  In the large-$N$
limit (\ref{ln}), the combination of $\alpha$, or equivalently, $a$, and
$N$ that remains finite is
\beq
\zeta \equiv \lim_{N \to \infty} \alpha \, N \ . 
\label{zeta_largeN}
\eeq
Correspondingly, the rescaled beta function that is finite has the form 
\beq
\beta_\zeta \equiv \frac{d\zeta}{dt} \ . 
\label{betazeta}
\eeq
where, as in Eq. (\ref{betaa}), $t=\ln \mu$. In this limit, for physical $p \in
(I_p)_{Adj}$, the (rescaled, finite) $\beta_{\zeta,2\ell}$, has an IR zero at
\beq
\zeta_{_{IR,2\ell},Adj} = \frac{8\pi(9-2p)}{32p-39} \ . 
\label{zetair_2loop_adjmodel_largeN}
\eeq
The approach to this limit of $N \to \infty$ involves correction terms that
are powers in $1/N$:
\beqs
N \alpha_{_{IR,2\ell,Adj}} = \frac{8\pi(9-2p)}{32p-39} - 
\frac{96\pi(p+48)}{(32p-39)^2 N} + 
O \bigg ( \frac{1}{N^2} \bigg ) \ .  \cr\cr
& & 
\label{zetair_2loop_adjmodel_largeNtaylor}
\eeqs
One may compare the approach to the $N \to \infty$ limit here with that in a
(vectorial) SU($N$) gauge theory with $N_f$ fermions in the fundamental
representation in the limit $N \to \infty$, $N_f \to \infty$ with the ratio
$N_f/N$ fixed and finite (and $\alpha(\mu)N$ a finite function of $\mu$),
denoted the LNN limit in \cite{lnn}. In that case \cite{bvh,lnn} the leading
correction term to the limit was suppressed like $1/N^2$ instead of $1/N$, and
the correction terms formed a series in powers of $1/N^2$ instead of powers in
$1/N$.  Hence, the approach to the $N \to \infty$ limit here is not as rapid as
in the LNN limit.


\subsection{Analysis of UV to IR Flows} 
\label{adjmodel_flows}

Because of the asymptotic freedom of the theory, i.e., the fact that the beta
function is negative for small $\alpha$, it follows that, as the Euclidean
reference momentum scale $\mu$ decreases from the ultraviolet toward the
infrared, $\alpha(\mu)$ increases.  There are several possibilities for the
behavior that can occur:

\begin{enumerate}

\item 

First, if the beta function has an IR zero at a sufficiently small value of
$\alpha = \alpha_{_{IR}}$, then one expects that the theory will evolve into
the infrared without any spontaneous chiral symmetry breaking.  In this case,
the IR zero of $\beta_\alpha$ is an exact IRFP of the renormalization group, so
that as $\mu \to 0$, the theory exhibits scale invariance with nonzero
anomalous dimensions.  In the IR limit $\mu \to 0$, one anticipates that the
theory is in a deconfined, massless non-Abelian Coulomb phase.

\item 

For smaller values of $p$, the IR zero of the beta function is larger, and
correspondingly, $\alpha(\mu)$ becomes larger as $\mu$ decreases from the UV to
the IR.  Then the strongly coupled gauge interaction can produce fermion
condensates that break global and possibly also local gauge symmetries. This
behavior also applies if $p$ is sufficiently small that the beta function has
no IR zero, so that $\alpha(\mu)$ keeps increasing with decreasing $\mu$ until
it exceeds the interval where the perturbative beta function describes its
evolution.  In this general category of UV to IR evolution, there can be a
sequence of condensate formations at various energy scales.

\item 

In the strongly coupled case (including both the subcases where the beta
function has an IR zero at sufficiently large coupling and where the beta
function has no IR zero), an alternate possibility is, if the fermion content
satisfies the 't Hooft anomaly-matching conditions \cite{thooft1979}, then the
gauge interaction might confine and produce massless gauge-singlet
composite fermions.

\end{enumerate} 

The beta function describes the growth of $\alpha(\mu)$ as the reference
momentum scale $\mu$ decreases from the UV to the IR.  If the fermion content
is such that the beta function has no IR zero, then the interaction definitely
becomes strongly coupled in the infrared.  If, on the other hand, the beta
function does have an IR zero, then one must investigate how large the value of
the coupling is at this zero.  In conjunction with knowledge of the probable
channel in which fermions may condense and the corresponding estimate of the
minimum critical coupling, $\alpha_{cr}$ that triggers this condensation, one 
can then draw a plausible inference as to whether the condensation takes place
or whether, in contrast, the theory evolves into the infrared without any
fermion condensation or associated spontaneous chiral symmetry breaking.  

The only composite fermions that one can form are those of the $p=0$ theory,
and we find that these do not match the global anomalies of $G_{fl,R{sc}}$
(given below in Eq. (\ref{gflcl_rselfconjugate}) for $R_{sc}=Adj$.  This rules
out the possibility that the original theory can form massless composite
fermions involving the full set of massless fermions in the theory with $p >
0$. As we will discuss below, however, if the UV to IR evolution leads to
sufficiently strong coupling so that there is condensation in the $R_{sc}
\times R_{sc} \to 1$ channel, giving the $R_{sc}$ fermions dynamical masses,
then in the low-energy effective field theory below the condensation scale,
with these fermions removed, the descendant theory is equivalent to the
original theory with $p=0$.  In this descendant theory (called the $S\bar F$
theory below), further evolution into the infrared might produce massless
gauge-singlet composite fermions.

To obtain information concerning the likely type of UV to IR evolution among
types 1 and 2 in the list above, as a function of $p$, we first identify the
most attractive channel, which is
\beq
Adj \times Adj \to 1 \ .
\label{adjadjto1_adjmodel}
\eeq
This clearly preserves the SU($N$) gauge symmetry, and has 
attractiveness measure
\beq
\Delta C_2 = 2N \quad {\rm for} \ Adj \times Adj \to 1 \ . 
\label{deltac2_adjadjto1_adjmodel}
\eeq
In particular, this channel is more attractive than the $S \times \bar F \to F$
channel, in accordance with the inequality (\ref{deltac2_inequality}). 
Quantitatively, the difference in $\Delta C_2$ values for these 
two channels is 
\beqs
& & \Delta C_2(Adj \times Adj \to 1)-\Delta C_2(S \times \bar F \to F) 
\cr\cr 
& = & \frac{N^2-N+2}{N} \ ,
\label{diffdeltaC2adjadj_sfbar}
\eeqs
which is positive for all physical $N$. The condensates for the $Adj \times Adj
\to 1$ channel are 
\beq
\langle \xi^{a \ T}_{b,i,L} C \, \xi^b_{a,j,L} \rangle \ , 
\quad i, \ j=1,...,p \ .  
\label{xixicondensate_adjmodel}
\eeq

From Eq. (\ref{deltac2_adjadjto1_adjmodel}), we obtain the rough estimate of
the minimal critical coupling for condensation in the $Adj \times Adj \to 1$
channel:
\beq
\alpha_{cr, Adj \times Adj} \simeq \frac{\pi}{3N} \ . 
\label{alfcrit_adjmodel}
\eeq
Thus, an approximate indication of the size of the IR fixed point relative to
the size that would lead to the formation of fermion condensates in the channel
(\ref{adjadjto1_adjmodel}) is the ratio
\beqs
& & \rho_{_{IR,Adj \times Adj }} \equiv 
\frac{\alpha_{_{IR,2\ell,Adj}}}{\alpha_{cr, Adj \times Adj}} \cr\cr
& = & 
\frac{24N^2[(9-2p)N-6]}{(32p-39)N^3+90N^2-3N-36} \ . 
\label{alfalfcrit_ratio_adjmodel}
\eeqs
As $p$ decreases, $\alpha_{_{IR,2\ell}}$ increases.  Therefore, considering $N$
and $p$ as being extended from the non-negative integers to the non-negative
real numbers, one can calculate a rough estimate of the critical value of $p$,
denoted $p_{cr,Adj \times Adj}$, such that, as $p$ decreases through this
value, $\alpha_{_{IR,2\ell}}$ increases through the value $\alpha_{cr,Adj
\times Adj}$. 
This critical value of $p_{cr,Adj \times Adj}$ is thus obtained by setting 
$R_{IR,Adj \times Adj}=1$ and solving for $p$, yielding
\beq
p_{cr,Adj \times Adj} \simeq \frac{3(85N^3-78N^2+N+12)}{80N^3} \ . 
\label{pcr_adjmodel}
\eeq
This critical value $p_{cr,Adj}$ is a monotonically increasing function of $N$
for physical $N$, increasing from $67/30=2.23$ for $N=3$ and, as $N \to
\infty$, 
\beq
\lim_{N \to \infty} p_{cr,Adj \times Adj} = \frac{51}{16} = 3.1875 \ ,
\label{pcrit_adjmodel_Ninf}
\eeq
where the limit is approach from below as $N$ increases. 

We list values of the ratio $\rho_{_{IR,Adj \times Adj}}$ in Table
\ref{alfir_and_ratio_adjmodel} for several illustrative values of $N$ and $p$.
For all of the values of $N$ presented in this table, the respective values of
the ratio $\rho_{_{IR,Adj \times Adj}}$ for $p=4$ are much smaller than 1, so
that one can conclude that for $p=4$, the theory evolves from the UV to a
scale-invariant, non-Abelian Coulomb phase in the IR.  As is evident from Table
\ref{alfir_and_ratio_adjmodel}, for a given $N$, as $p$ decreases,
$\alpha_{_{IR,2\ell,Adj}}$ increases.  As this IR coupling becomes of O(1), the
uncertainties in the use of perturbation theory increase.  For most of $p=3$
cases shown with various $N$, the ratio $\rho_{_{IR,Adj \times Adj}}$ is
sufficiently close to 1 that, taking account of these uncertainties, one cannot
draw a definite conclusion as to whether fermion condensate does or does not
take place.  For the cases shown in Table \ref{alfir_and_ratio_adjmodel} with
$p=1$ (where this is in $(I_p)_{Adj}$) and $p=2$, the ratio $\rho_{_{IR,Adj
\times Adj}}$ is substantially larger than 1, so that in these cases, one
expects that the gauge interaction become strong enough to produce fermion
condensation in the channel (\ref{adjadjto1_adjmodel}).

In the large-$N$ limit defined above, 
\beq
\lim_{N \to \infty} \rho_{_{IR,Adj \times Adj}} = \frac{24(9-2p)}{32p-39} \ .
\label{alfalfcrit_ratio_adjmodel_largeN}
\eeq
In particular, 
\beq 
\lim_{N \to \infty} \rho_{_{IR, Adj \times Adj}}
=\frac{24}{89} = 0.270 \quad {\rm for} \ p=4
\label{alfalfcrit_ratio_adjmodel_largeN_peq4}
\eeq
\beq
\lim_{N \to \infty} \rho_{_{IR, Adj \times Adj}} = 
\frac{72}{57} = 1.26 \quad {\rm for} \ p=3
\label{alfalfcrit_ratio_adjmodel_largeN_peq3}
\eeq
\beq
\lim_{N \to \infty} \rho_{_{IR, Adj \times Adj}} = 
\frac{24}{5} = 4.80  \quad {\rm for} \ 
p=2 
\label{alfalfcrit_ratio_adjmodel_largeN_peq2}
\eeq
(where the floating-point results are given to the indicated accuracy).  Hence,
in this large-$N$ limit, since the limit of the ratio $\rho_{_{IR,Adj \times
Adj}}$ for $p=4$ is sufficiently small compared to 1 that it is plausible that
in the IR the theory is in a deconfined Coulombic phase, while if $p=3$,
$\rho_{_{IR,Adj \times Adj}}$ is too close to unity for one to be able to draw
a definite conclusion.  Finally, if $p=2$, then $\rho_{_{IR,Adj \times Adj}}$
is sufficiently large compared with 1 that one expects that the theory can
produce bilinear condensates in the most attractive channel, as discussed
above.

We continue with the analysis of the UV to IR evolution for the smaller values
of $p$ that produce a strongly coupled gauge interaction.  As the momentum
scale $\mu$ decreases through a scale denoted $\Lambda_{Adj}$, $\alpha(\mu)$
exceeds $\alpha_{cr,Adj}$, and, from our discussion above, we infer that the
gauge interaction produces the bilinear fermion condensates
(\ref{xixicondensate_adjmodel}) in the MAC, $Adj \times Adj \to 1$.  These
condensates preserve the SU($N$) gauge symmetry and the U(1)$_1$ global
symmetry, while breaking the U(1)$_2$ and SU($p$) global symmetries.  By the
use of a vacuum alignment argument \cite{vacalign}, one can plausibly infer
that the condensates (\ref{xixicondensate_adjmodel}) have $i=j$, with
$i=1,...,p$ and hence preserve an SO($p$) global isospin symmetry defined by
the transformation
\beq
\xi^a_{b,i,L} \to \sum_{j=1}^p {\cal O}_{ij} \xi^a_{b,j,L} \ , 
\quad {\cal O} \in {\rm SO}(p) \ . 
\label{sop}
\eeq
Just as light quarks gain dynamical, constituent quark masses of order
$\Lambda_{QCD}$ due to the formation of $\langle \bar q q \rangle$ condensates
in quantum chromodynamics (QCD), so also, the $p(N^2-1)$ components,
$\xi^a_{b,i,L}$, of the $Adj$ fermions involved in these condensates pick up a
common dynamical mass of order $\Lambda_{Adj}$.

At scales $\mu < \Lambda_{Adj}$, the analysis proceeds by integrating out the
massive $\xi^a_{b,j,L}$ fermions, constructing the low-energy effective field 
theory applicable for these lower scales, and then exploring the further
evolution of this descendant theory into the infrared.  Since the condensation 
(\ref{xixicondensate_adjmodel}) gives dynamical masses to all of the $Adj$
fermions $\xi^a_{b,j,L}$, $j=1,...,p$, the low-energy effective theory below
this condensation scale $\Lambda_{Adj}$ is just the $p=0$ theory.  Since the
evolution of this theory is the same as for our second type of chiral gauge
theory, we first study this second theory, and then discuss the further IR
evolution.  


\section{Theory with $N=2k$ and $R =[N/2]_N$}
\label{atmodel}


\subsection{Particle Content} 
\label{atmodel_content}

In this section we construct and study a chiral gauge theory with gauge group
$G={\rm SU}(N)$ with even $N=2k$, and fermions transforming according to
\begin{enumerate}

\item
a symmetric rank-2 tensor representation, $S$, with corresponding 
field $\psi^{ab}_L = \psi^{ba}_L$,

\item $N+4$ copies chiral fermions in the conjugate fundamental representation,
$\overline{\fund}$, with fields $\chi_{a,i,L}$, $i=1,...,N+4$, and

\item 
$p$ copies of chiral fermions in the totally antisymmetric $k$-fold tensor
  representation $[N/2]_N = [k]_{2k}$, with fields 
  $\xi^{a_1...a_k}_{j,L}$, $j=1,...,p$.

\end{enumerate}
We again label this theory by the representation of the fermions in the
vectorlike subsector, namely AT, for \underline{a}ntisymmetric $k$-fold
\underline{t}ensor. This fermion content is summarized in Table 
\ref{fermion_properties_rscmodels}. 

The representation $[k]_N$ has the dimension (for general $N$)
\beq
{\rm dim}([k]_N) = {N \choose k}
\label{dimkn}
\eeq
and satisfies the equivalence property 
\beq
[N-k]_N = \overline{[k]}_N \ . 
\label{kequivrel}
\eeq
Here we have used the standard notation for the binomial coefficient, ${a
\choose b} \equiv a!/[b! \, (a-b)!]$. An important property that follows from
Eq. (\ref{kequivrel}) that that we will use here is the fact that for our case
of interest, $N=2k$, the representation $[k]_{2k}$ is self-conjugate:
\beq
[k]_{2k} = \overline{[k]}_{2k} \ . 
\label{k2kselfconjugate}
\eeq
Combining the self-conjugate property of 
$[N/2]_N = [k]_{2k}$ with the relation (\ref{anomrrbar}), it follows that 
\beq
{\rm Anom}([k]_{2k}) = 0 \ . 
\label{k2k_noanom}
\eeq
Thus, this theory has the same irreducibly chiral sector as the theory 
discussed in the previous section, and a vectorlike subsector that consists
of the $p$ copies of the fermions in the $[N/2]_N$ representation.  


\subsection{Beta Function}
\label{atmodel_betafunction}

We calculate that the one- and two-loop terms in the beta function of this 
theory are, in terms of $k=N/2$, 
\beq
(b_1)_{AT} = 6k-2-\frac{p \, (2k-2)!}{3[(k-1)!]^2}
\label{b1_atmodel}
\eeq
and 
\beq
(b_2)_{AT} = \frac{52k^3-60k^2+k+6}{2k}- 
\frac{p \, k(43+6k)\, (2k-2)!}{12[(k-1)!]^2} \ .
\label{b2_atmodel}
\eeq
For small $p$, $(b_1)_{AT}$ is positive, and as $p$ increases, 
$(b_1)_{AT}$ decreases and passes through zero as $p$ exceeds the value
\beq
p_{b1z,AT} =  \frac{6(3k-1)[(k-1)!]^2}{(2k-2)!} \ . 
\label{pb1z_atmodel}
\eeq
The requirement that the theory should be asymptotically free is thus satisfied
if
\beq
p <  \frac{6(3k-1)[(k-1)!]^2}{(2k-2)!} \ . 
\label{p_upperbound_atmodel}
\eeq
This upper bound decreases rapidly as a function of $k=N/2$, so that as $k$
increases, eventually the requirement of asymptotic freedom precludes any
nonzero value of $p$. Thus, the AT theory has no asymptotically free large-$N$
limit with nonzero $p$, in contrast to the $Adj$ and $S \bar S$ theories
constructed and studied here and the $Sp$ theory reviewed in Sect. 
\ref{spmodel}.

The beta function of the AT theory has an IR zero if $b_2$ is negative.  For
small $p$, $(b_2)_{AT}$ is positive, and it decreases through zero to negative
values as $p$ (continued to the real numbers) increases through the value
\beq
p_{b2z,AT} = \frac{6(52k^3-60k^2+k+6) \, [(k-1)!]^2}{k^2(6k+43) \, (2k-2)!} 
\ . 
\label{pb2z_atmodel}
\eeq
We observe that $p_{b1z,AT} >  p_{b2z,AT}$.  This is proved by considering the
difference, 
\beqs
& & p_{b1z,AT} - p_{b2z,AT} \cr\cr
& = & \frac{6(18k^4+71k^3+17k^2-k-6)[(k-1)!]^2}{k^2(43+6k)[(2k-2)!]} \ . 
\cr\cr
& & 
\label{pb1z_atmodel_minus_pb2z_atmodel}
\eeqs
This difference is positive for all $k$ values of relevance here (with $k$
extended to the positive reals, it is positive for $k > 0.3724$).  By itself,
this inequality does not guarantee that there is an integral value of $p$ that
lies above $p_{b2z,AT}$ and below $p_{b1z,AT}$, but in fact we find that for
each relevant case, there are one or more such integral values.  These then
define the respective intervals $(I_p)_{AT}$, 
\beq
(I_p)_{AT}: \quad p_{b2z,AT} < p < p_{b1z,AT}
\label{pinterval_atmodel}
\eeq
for each $k$. For the (integral) values of $p \in (I_p)_{AT}$, the beta
function of the SU($2k$) AT theory has an IR zero.  We list the values of
$p_{b1z}$, $p_{b2z}$, $p_{max}$, and $(I_p)_{AT}$ in Table
\ref{pb12z_atmodel}).  Note that for the cases $G={\rm SU}(N)$ with $k \ge 2$
under consideration here, the requirement of asymptotic freedom allows nonzero
values of $p$ only for $k \le 5$.

For a given $N=2k$ whith a nonvacuous interval $(I_p)_{AT}$, the
$\beta_{\alpha,2\ell}$ has an IR zero at 
\beq
\alpha_{_{IR,2\ell,AT}} = -\frac{4\pi (b_1)_{AT}}{(b_2)_{AT}} 
\label{alfir_2loop_atmodel}
\eeq
where $(b_1)_{AT}$ and $(b_2)_{AT}$ were given in Eqs. (\ref{b1_atmodel}) and
(\ref{b2_atmodel}) above.  We list the values of $\alpha_{_{IR,2\ell,AT}}$ in
Table \ref{alfir_and_ratio_atmodel}. 


\subsection{UV to IR Evolution} 

Here we analyze the UV to IR evolution of this AT chiral gauge theory.  By 
construction, the most attractive channel involves fermion condensation in the
channel (\ref{rrbarto1_channel}), with $R=[N/2]_N = [k]_{2k}$ in this case, 
i.e., 
\beq
[N/2]_N \times [N/2]_N \to 1 \ . 
\label{kkchannel}
\eeq
This preserves the SU($N$) gauge symmetry and has the attractiveness measure 
\beq
\Delta C_2 = 2C_2([N/2]_N) = \frac{k(2k+1)}{2} \ ,
\label{deltac2_kkchannel}
\eeq
where we have used the result for $C_2([k]_N)$ given in
Appendix \ref{bn}.  The condensates are
\beq
\langle \epsilon_{a_1,...a_{2k}} \, \xi^{a_1,...,a_k \ T}_{i,L} C \,
\xi^{a_{k+1},...,a_{2k}}_{j,L} \rangle \ , \ i,j = 1,..., p \ . 
\label{kkcondensate}
\eeq
By a vacuum alignment argument, one may infer that these condensates have $i=j$
\cite{vacalign}.  To show that the channel (\ref{kkchannel}) is more attractive
than the next-most-attractive channel, $S \times \bar F \to F$, we examine the
difference
\beqs
& & \Delta C_2([N/2]_N \times [N/2]_N \to 1) - 
\Delta C_2(S \times \bar F \to F) \cr\cr
& = & 2C_2([N/2]_N) - C_2(S) = \frac{2k^3-3k^2-2k+2}{2k} \ . 
\cr\cr
& & 
\label{dc2difkk}
\eeqs
This difference is positive for all values of $k \ge 2$ of interest here. 

If the beta function has no IR zero, then as the scale $\mu$ decreases and
$\alpha(\mu)$ increases, it will eventually become large enough to cause
condensation, which, according to the MAC criterion, will be in this channel
(\ref{kkchannel}).  If the beta function does have a zero, then the next step
in the analysis is to determine how the value of the coupling at this zero
compares with $\alpha_{cr}$ for the most attractive channel,
(\ref{kkchannel}).  Substituting (\ref{deltac2_kkchannel}) into 
the general formula for Eq. (\ref{alfcrit}), we calculate 
\beq
\alpha_{cr,AT} = \frac{4\pi}{3k(2k+1)} \ . 
\label{alfcrit_kkchannel}
\eeq
As discussed above, an approximate measure of how strong the coupling gets in
the infrared, compared with the minimum critical value for condensation in the
MAC is then given by the ratio
\beq
\rho_{_{IR,AT}} \equiv \frac{\alpha_{_{IR,2\ell,AT}}}{\alpha_{cr,AT}} 
\ .
\label{ratiok2k}
\eeq
We list values of $\rho_{_{IR,AT}}$ for the relevant $N$ and $p$ in Table
\ref{alfir_and_ratio_atmodel}.  In cases where condensation occurs in this
theory we denote the scale at which it occurs as $\Lambda_{[N/2]_N}$.


\subsubsection{AT Theory with $G={\rm SU}(4)$}

In this subsection and the following ones we discuss three illustrative cases
with various values of $N=2k$ and their corresponding intervals $(I_p)_{AT}$.
For each value of $N$, if $p$ is nonzero and $p < p_{b2z}$, i.e., below the
lower end of the interval $(I_p)_{AT}$, then the theory has no IR fixed point,
even an approximate one, so that the gauge coupling continues to grow in the
infrared and will cause condensation in the MAC.  Hence, we restrict our
consideration here to $p \in (I_p)_{AT}$.  The reader is referred to Tables
\ref{pb12z_atmodel} and \ref{alfir_and_ratio_atmodel} for numerical values
of relevant quantities.  As indicated in Table \ref{pb12z_atmodel}, for
this SU(4) AT theory the interval $(I_p)_{AT}$ is $3 \le p \le 14$.  For $p$ in
this interval, $\beta_{\alpha,2\ell,AT}$ has an IR zero at
\beq
N=4: \ \alpha_{_{IR,2\ell,AT}} = \frac{8\pi(15-p)}{55p-138}  \ . 
\label{air_2loop_keq2_atmodel}
\eeq
The ratio $\rho_{_{IR,AT}}$ is 
\beq
N=4: \ \rho_{_{IR,AT}} = \frac{60(15-p)}{55p-138} \ . 
\label{ratio_kasym}
\eeq
As listed in Table \ref{alfir_and_ratio_atmodel}, for the range of $p$ from 3
to 7, this ratio takes on values decreasing from 26.7 to 1.94, all well above
unity. Thus, one may plausibly expect that for these values of $p$, in the UV
to IR evolution, as the reference scale $\mu$ decreases sufficiently and the
running coupling approaches $\alpha_{_{IR,2\ell,AT}}$, the gauge interaction
will become strong enough to cause fermion condensation in the most attractive
channel, $[2]_4 \times [2]_4 \to 1$.  For $p=8, \ 9, \ 10, \ 11$,
$\rho_{_{IR,AT}}$ has the respective values 1.39, \ 1.01, \ 0.728, 0.514.
Given the theoretical uncertainties in these estimates, the IR behavior might
or might not involve the formation of the condensates (\ref{kkcondensate}).
For the largest values of $p$, namely $p=12, \ 13, \ 14$, $\rho_{_{IR,AT}}$ has
the respective values 0.345, \ 0.208, \ 0.095, so for these cases, it is likely
that the theory evolves from the UV to a scale-invariant, deconfined, Coulombic
IR phase. This inference is, of course, most reliable for the largest allowed
value of $p$, namely $p=14$, which leads to the smallest value of
$\alpha_{_{IR,2\ell,AT}}$ and $\rho_{_{IR,AT}}$.  As discussed above, in the
cases where there is condensate formation and chiral symmetry breaking, the
IRFP is only approximate, while in the cases where there is no such chiral
symmetry breaking the IRFP (calculated to all orders) is exact.


\subsubsection{AT Theory with $G={\rm SU}(6)$}

In the SU(6) \ (i.e., $k=3$) AT theory, $(I_p)_{AT}$ is the interval 
$2 \le p \le 7$.  For $p$ in this interval, $\beta_{\alpha,2\ell}$ 
has an IR zero at
\beq
N=6: \ \alpha_{_{IR,2\ell,AT}} = \frac{16\pi(8-p)}{3(61p-97)} \ . 
\label{alfir_2loop_keq3}
\eeq
The ratio $\rho_{_{IR,AT}}$ is 
\beq
N=6: \ \rho_{_{IR,AT}} = \frac{84(8-p)}{61p-97} \ . 
\label{ratio_keq3}
\eeq
As listed in Table \ref{alfir_and_ratio_atmodel}, for 
$2 \le p \le 7$, this has the respective values 20.16, \ 4.89, \ 2.29, \
1.21, \ 0.625, \ 0.255.  Thus, for $p=3$ and $p=4$, it is likely that
condensation occurs in the MAC, $[3]_6 \times [3]_6 \to 1$ channel; for $p=7$,
it is likely that there is no condensation; and for the middle two values $p=5$
and $p=6$, taking account of the intrinsic theoretical uncertainties, one
cannot give a very definite prediction from this analysis. 


\subsubsection{AT Theory with $G={\rm SU}(10)$}
\label{su10}

In the SU(10) \ ($k=5$) AT theory, the interval $(I_p)_{AT}$ reduces to just a
single nonzero value, $p=1$, and the resultant $\alpha_{_{IR,2\ell,AT}} =
0.036$, yielding the ratio $\rho_{_{IR,AT}}= 0.473$. It is thus likely that
this theory evolves from the UV to the IR to a non-Abelian Coulomb phase,
although there are obvious uncertainties in this inference due to the
strong-coupling physics involved.


\section{Global Flavor Symmetry for Theories with Self-Conjugate $R$} 
\label{global_flavor_symmetry}

In analyzing the global flavor symmetry of these chiral gauge
theories, it is useful to consider a more general class of theories, in which
the vectorlike fermion subsector is comprised of fermions transforming under a
general self-conjugate representation, $R=R_{sc}$. The results will then
be applied to the two specific theories discussed above, namely those with
$G={\rm SU}(N)$, $N \ge 3$, and $R_{sc}=Adj$; and the AT theory with
$G={\rm SU}(N)$ with even $N=2k$, $k \ge 2$, and $R_{sc}=[N/2]_N$. 

The classical global chiral flavor symmetry of a theory in this class of
theories is
\beqs
& & G_{fl,cl,R_{sc}} = {\rm U}(1)_S \otimes {\rm U}(N+4)_{\bar F} \otimes 
{\rm U}(p)_{R_{sc}} \cr\cr
& = & {\rm U}(1)_S \otimes {\rm SU}(N+4)_{\bar F} \otimes 
{\rm U}(1)_{\bar F} \otimes {\rm SU}(p)_{R_{sc}} \otimes {\rm U}(1)_{R_{sc}}
 \ . \cr\cr
& &
\label{gflcl_rselfconjugate}
\eeqs
The representations of the fermions in the two theories with $R=R_{sc}$ under 
this symmetry are given in Table \ref{fermion_properties_rscmodels}. 
The corresponding global unitary transformations are
\beq
\psi^{ab}_L \to U_S \, \psi^{ab}_L \ , \quad U_S \in {\rm U}(1)_S \ , 
\label{us}
\eeq
\beq
\chi_{a,i,L} \to \sum_{j=1}^{N+4}(U_{\bar F})_{ij} \, \chi_{a,j,L} \ , 
\quad U_{\bar F} \in {\rm U}(N+4)_{\bar F} \ , 
\label{ufbar}
\eeq
and 
\beq
\xi_{i,L} \to \sum_{j=1}^p (U_{R_{sc}})_{ij} \, \xi_{j,L} \ , 
\quad U_{R_{sc}} \in {\rm U}(p)_{R_{sc}}
\label{uxi}
\eeq
where we have suppressed the SU($N$) gauge indices in Eq. (\ref{uxi}), which 
applies to each theory with the corresponding $\xi$ field, i.e., 
$\xi^a_{b,i,L}$ in the $Adj$ theory and $\xi^{a_1,...,a_k}_{i,L}$ in the AT
theory. 

Each of the three global U(1) symmetries is broken the instantons of 
the SU($N$) gauge theory \cite{adjanom}.  One
may define a three-dimensional vector of anomaly factors, 
\beqs
{\vec v} & = & \Big ( N_S T(S), \ N_{\bar F} T(\bar F), \ 
N_{R_{sc}} T(R_{sc}) \Big ) \cr\cr
     & = & \bigg ( \frac{N+2}{2}, \ \frac{N+4}{2}, \ p T_{R_{sc}} \Big ) \ , 
\label{anomvec}
\eeqs
where the basis is $(S,\bar F,R_{sc})$, and we have inserted the values
$N_S=1$, $N_{\bar F}=N+4$, and $N_{R_{sc}}=p$.  One can construct two linear
combinations of the three original currents that are conserved in the presence
of SU($N$) instantons.  The fermions have charges under these global U(1)
symmetries given by the vectors 
\beq
{\vec Q}^{(j)} \equiv \Big ( Q^{(j)}_S, \ Q^{(j)}_{\bar F}, 
Q^{(j)}_{R_{sc}} \Big ), \quad j=1,2 \ , 
\label{qvec}
\eeq
where $j=1$ for U(1)$_1$ and $j=2$ for U(1)$_2$.  The condition that the
corresponding currents are conserved, i.e., the U(1)$_j$ global 
symmetries are exact, in the presence of instantons is that 
\beq
\sum_f N_f T(R_f) \, Q_f^{(j)} = {\vec v} \cdot {\vec Q}^{(j)} = 0 
\quad {\rm for} \ j=1, \ 2 \ . 
\label{u1inv}
\eeq
As indicated, this condition is equivalent to the condition that the vectors of
charges under the U(1)$_1$ and U(1)$_2$ symmetries are orthogonal to the vector
${\vec v}$. (Note that the condition (\ref{u1inv}) does not uniquely determine
the vectors ${\vec Q}^{(j)}$, $j=1,\ 2$.  It will be convenient to choose the
first vector, ${\vec Q}^{(1)}$, so that $Q_{R_{sc}}^{(1)}=0$.  We thus choose
\beq
{\vec Q}^{(1)} = \Big ( N+4, \ -(N+2), \ 0 \Big ) \ . 
\label{qvec1_rselfconjugate}
\eeq
For the vector of charges under U(1)$_2$, we choose
\beq
{\vec Q}^{(2)} = \Big ( 2p T_{R_{sc}}, \ 0, \ -(N+2) \Big ) \ . 
\label{qvec2_rselfconjugate}
\eeq
(Note that in contrast to Gram-Schmidt orthogonalization of the 
three vectors ${\vec v}$, ${\vec Q}^{(1)}$, and ${\vec Q}^{(2)}$, here
it is not necessary that ${\vec Q}^{(1)} \cdot {\vec Q}^{(2)} = 0$.) 

The actual non-anomalous global chiral flavor symmetry group of the class
of chiral gauge theories with $R=R_{sc}$ is then 
\beqs
G_{fl,R_{sc}} = {\rm SU}(N+4)_{\bar F} \otimes {\rm SU}(p)_{R_{sc}} \otimes 
{\rm U}(1)_1 \otimes {\rm U}(1)_2 \ . 
\cr\cr
& & 
\label{gfl_rselfconjugate}
\eeqs
For the two respective theories with (i) $R_{sc}=Adj$ and 
(ii) $R_{sc}=[N/2]_N$, 
Eqs. (\ref{qvec2_rselfconjugate}) and (\ref{gfl_rselfconjugate}) apply with
(i) $T_{R_{sc}} = T(Adj)=N$ and (ii) $T_{[N/2]_N}$ given by Eq. (\ref{tk2k}) in
Appendix \ref{bn}. We summarize these properties in Table
\ref{fermion_properties_rscmodels}.

In general, one must also check to see if either of the chiral gauge theories 
with $R_{sc}=Adj$ or $R_{sc}=[N/2]_N$ satisfies the 't Hooft anomaly-matching
conditions, which are necessary conditions for the possible formation of
massless gauge-singlet composite fermions.  The possible gauge-singlet fermions
can be described by wavefunctions of the form 
\beq
B_{ij} = \bar F_{a,i,L} \, S^{ab}_L \, \bar F_{b,j,L}  \ , 1 \le i,j \le N+4 
\ . 
\label{bij}
\eeq
Given the minus sign from Fermi statistics and the fact that $S^{ab}$ is a
rank-2 symmetric tensor representation ($\sym$) of SU($N$), it follows that
$B_{ij} =-B_{ji}$, i.e., $B_{jk}$ is a rank-2 antisymmetric tensor
representation ($\asym$) of the ${\rm SU}(N+4)_{\bar F}$ factor group in the
global flavor symmetry group $G_{fl}$. There are thus $(N+4)(N+3)/2$ components
of $B_{ij}$.  The charges of $B_{ij}$ under the two global abelian factor
groups in $G_{fl,R_{sc}}$, U(1)$_k$, $k=1,2$ are determined by the relation
\beq
Q_B^{(k)} = Q_S^{(j)} + 2Q_{\bar F}^{(k)} \ , \quad k=1,2 
\label{qb}
\eeq
Hence, 
\beq
Q_B^{(1)} = -N 
\label{qb_u1_rsc}
\eeq
and
\beq
Q_B^{(2)} = 2p \, T_{R_{sc}} \ . 
\label{qb_u2_rsc}
\eeq
We find that the global anomalies of a theory with these massless composite 
fermions do not match those of the original $G_{fl}$ group except in the
degenerate case $p=0$.  This $p=0$ case describes a descendant low-energy
effective field theory that occurs if there is condensation in the $R_{sc}
\times R_{sc} \to 1$ channel, and will be discussed below. 


\section{Analysis of Low-Energy Effective Theory for $\mu < \Lambda_{R_{sc}}$} 
\label{leeft}

In the cases where the values of $N$ and $p$ are such as to lead to the
respective bilinear fermion condensates (\ref{xixicondensate_adjmodel}) or 
(\ref{kkcondensate}) at the corresponding scales $\Lambda_{Adj}$ or
$\Lambda_{[N/2]_N}$, we analyze the further UV to IR evolution below these
scales.  We denote these scales generically as $\Lambda_{R_{sc}}$.  Because of
this condensation, the $p$ fermions $\xi^a_{b,i,L}$ involved in the condensate
(\ref{xixicondensate_adjmodel}) in the $Adj$ model and the $p$ fermions 
$\xi^{a_1,...,a_l}_{i,L}$ involved in the condensate (\ref{kkcondensate}) in
the AT theory gain dynamical masses of order $\Lambda_{Adj}$ and 
$\Lambda_{[N/2]_N}$, respectively. 

For momentum scales $\mu$ slightly below the condensation scale 
$\Lambda_{R_{sc}}$, the resultant global symmetry is 
\beq
G_{fl}' = {\rm SU}(N+4)_{\bar F} \otimes {\rm SO}(p) \otimes {\rm U}(1)_1 \ . 
\label{gflprime_rsc} 
\eeq
Here the ${\rm SU}(N+4)_{\bar F} \otimes {\rm U}(1)_1$ is a global chiral
symmetry operating on the massless $S$ and $\bar F$ fermions, leaving their
covariant derivatives invariant, while the SO($p$)
is a global isospin symmetry of the condensate in each of our two theories with
$R=R_{sc}$, or equivalently, the corresponding effective mass term.  
These mass terms are
\beq
\Lambda_{Adj} \, \sum_{i=1}^p \ \xi^{a \ T}_{b,i,L} C \, \xi^b_{a,i,L} + h.c. 
\label{adjmass}
\eeq
in the $Adj$ theory and 
\beq
\Lambda_{[N/2]_N} \, \sum_{i=1}^p 
\langle \epsilon_{a_1,...a_{2k}} \, \xi^{a_1,...,a_k \ T}_{i,L} C \,
\xi^{a_{k+1},...,a_{2k}}_{i,L} \rangle + h.c.
\label{k2kmass}
\eeq
in the AT theory produced by the bilinear fermion condensations in 
these respective theories. This SO($p$) symmetry also leaves the covariant
derivatives of these $\xi$ fields invariant. 

The spontaneous symmetry breaking of the initial nonanomalous global symmetry
$G_{fl}$ in Eq.  (\ref{gfl_rselfconjugate}) to the final global symmetry
(\ref{gflprime_rsc}) produces
\beq
o({\rm SU}(p))+1-o({\rm SO}(p)) = \frac{p(p+1)}{2} \ . 
\label{firstngb_rsc}
\eeq
massless Nambu-Goldstone bosons, where $o(H)$ denotes the order of a group $H$.

As the reference scale $\mu$ decreases well below $\Lambda_{R_{sc}}$, we
integrate these now-massive $\xi$ fermions out of the low-energy (LE) effective
field theory (LEEFT) applicable for $\mu \ll \Lambda_{R_{sc}}$.  Focusing on
the infrared region $\mu \ll \Lambda_{R_{sc}}$, with the $\xi$ fermions
integrated out, both the theory with $R_{sc}=Adj$ and the theory with 
$R_{sc}=[N/2]_N$ reduce to the same low-energy descendant theory, with 
(massless) $S$ fermion and $N+4$ copies of $\bar F$ fermions.
We denote this as the $S\bar F$ theory.  This theory has been well studied in
the past \cite{rds_masslessfermions,mac,by,dfcgt,ads,cgt}.  We recall the
results from these earlier studies that we will need for our present analysis.

The value of $f_{UV}$ for the $S \bar F$ model, which we denote as
$f_{UV,S\bar F M}$ ($M$ standing for model), is given by the $p=0$ special case
of Eq. (\ref{fuv_rsc}), namely
\beqs
& & f_{UV,S \bar F M} = 2(N^2-1)+\frac{7}{4}\Big [ \frac{N(N+1)}{2}+(N+4)N 
\Big ] \ . \cr\cr
& & 
\label{fuv_sfbar}
\eeqs
The $S\bar F$ theory is invariant under a nonanomalous global flavor symmetry
group
\beq
G_{fl,S\bar F M} = {\rm SU}(N+4)_{\bar F} \otimes {\rm U}(1)_{S \bar F}  \ . 
\label{gfle}
\eeq
For this theory the three-dimensional vector (\ref{anomvec}) reduces to a
two-dimensional vector with the third entry deleted, and the vector of charges
that is orthogonal to it and hence defines the charge assignments of the 
U(1)$_{S \bar F}$ is given by the first two entries in $Q^{(1)}$, namely 
\beq
{\vec Q}^{(1)} = (N+4,-(N+2)) \ . 
\label{q1leeft}
\eeq

The $S\bar F$ theory is asymptotically free, so the gauge coupling continues to
grow as $\mu$ decreases.  The beta function of this $S\bar F$ theory has
one-loop and two-loop coefficients given by Eqs. (\ref{b1_adjmodel}) and
(\ref{b2_adjmodel}) with $p=0$ or equivalently, the $p=0$ special case of
Eqs. (\ref{b1_spmodel}) and (\ref{b2_spmodel}). In the relevant range $N \ge
3$, $b_2$ is positive.  Since $b_1$ and $b_2$ thus have the same sign, the beta
function, calculated to the maximal scheme-independent order of two loops, does
not have any IR zero.  Hence, as $\mu$ decreases from the UV to the IR, the
running coupling $\alpha(\mu)$ increases, eventually exceeding the region where
the perturbatively calculated beta function is applicable.

There are two possible types of UV to IR evolution in the $S\bar F$ theory.
First, the strongly coupled gauge interaction may produce bilinear fermion
condensates. The most attractive channel is $S \times \bar F \to F$, with
condensates
\beq
\langle \sum_{b=1}^N \psi^{ab \ T}_L C \chi_{b,i,L}\rangle \ . 
\label{sfbar_condensate}
\eeq
Without loss of generality, one may take $a=N$ and $i=1$ for the first
condensate.  This breaks the SU($N$) gauge symmetry down to SU($N-1$), so that
the $2N-1$ gauge bosons in the coset ${\rm SU}(N)/{\rm SU}(N-1)$ gain masses of
order this scale of condensation, which we denote $\Lambda_N$.  The fermions
$\psi^{Nb}_L$ and $\chi_{b,1,L}$ with $b=1,..,N$ involved in this condensate
also gain dynamical masses of order $\Lambda_N$.  In the low-energy theory
applicable for scales $\mu < \Lambda_N$, these now massive fermions are
integrated out.

The descendant theory is again asymptotically free, so the
gauge coupling inherited from the SU($N$) theory continues to increase.  There
is then a second condensation, again in the MAC, $S \times \bar F \to F$ 
channel, breaking the gauge symmetry from SU($N-1$) to SU($N-2$). Without loss
of generality, we may take the breaking direction to be $a=N-1$ and the $\bar
F$ fermion involved in the condensate to be labelled as $\chi_{b,2,L}$, so that
this condensate is 
\beq
\langle \sum_{b=1}^{N-1} \psi^{N-1,b \ T}_L C \chi_{b,2,L}\rangle \ . 
\label{sfbar_condensate2}
\eeq
We denote the scale at which this occurs as $\Lambda_{N-1}$.  The $2N-3$ gauge
bosons in the coset ${\rm SU}(N-1)/{\rm SU}(N-2)$ gain masses of order
$\Lambda_{N-1}$ and the fermions $S^{N-1,b}_L$ and $\chi_{b,2,L}$ with 
$b=1,...,N-1$ involved in this condensate gain dynamical masses of order
$\Lambda_{N-1}$.  This sequential breaking via condensation in the respective
$S \times \bar F \to F$ channels continues at the scales $\Lambda_{N-2}$,
etc. until the gauge symmetry is completely broken. Thus, the sequence of gauge
symmetry breaking is 
\beq
{\rm SU}(N) \to {\rm SU}(N-1) \to \cdots \to SU(2) \to \emptyset \ . 
\label{sequentialbreaking}
\eeq
The gauge bosons in the respective cosets ${\rm SU}(N)/{\rm SU}(N-1)$, ${\rm
SU}(N-1)/{\rm SU}(N-2)$, etc. gain masses of order $\Lambda_N$,
$\Lambda_{N-1}$, etc, as do the components of the fermions involved in the
respective condensates.  

Considering the $S\bar F$ theory, for
this type of UV to IR evolution \cite{dfcgt,ads,cgt}, 
\beq
f_{IR,S \bar F M; S \times \bar F} 
= 8N+1+\frac{7}{4}\Big [ \frac{N(N-1)}{2} + 4N \Big ] \ , 
\label{firbk_pzero}
\eeq
where here the subscript $S \bar F M$ means the $S \bar F$ model, and the
subscript $S \bar F$ refers to the condensation channel. 
For the $S\bar F$ model, with this type of UV to IR evolution, one
then has
\beqs
(\Delta f)_{S\bar F M; S \times \bar F} & = &
f_{UV,S\bar F M} - f_{IR, S \bar F M; S \times \bar F} \cr\cr
& = & \frac{15N^2-25N-12}{4} \ . 
\label{deltaf_sfbar_bk}
\eeqs
This is positive for all relevant values of $N$.  (For $N$ extended to the
positive reals, it is positive for $N > (25+\sqrt{1345} \ )/30 = 2.056$.) 

The low-energy effective $S\bar F$ theory applicable below $\Lambda_{R_{sc}}$
could also undergo a different type of flow deeper into the infrared, namely
one leading to confinement with massless gauge-singlet composite fermions with
wavefunctions (\ref{bij}). In this case, for this
$S \bar F$ theory, considered in isolation, 
\beq
f_{IR,S\bar F M;sym} = 
\frac{7}{4}\Big [ \frac{(N+4)(N+3)}{2} \Big ]  \ . 
\label{firsym_pzero}
\eeq
Hence, for this type of UV to IR evolution, 
\beq
(\Delta f)_{S\bar F M;sym} = \frac{15N^2+7N-50}{4} \ . 
\label{deltaf_sfbar_sym}
\eeq
This is positive for all relevant values of $N$. (For $N$ extended to the
positive reals, it is positive for $N > (-7+\sqrt{3049} \ )/30 = 1.607$.) 
Thus, for both of these types of UV to IR evolution of the $S \bar F$ theory,
the conjectured degree-of-freedom inequality (\ref{dfi}) is obeyed. 


\section{Comparison with Degree-of-Freedom Inequality}
\label{dfi_comparison}

We now combine the results for the $S\bar F$ theory with our calculations of UV
and IR degree-of-freedom counts for the different types of UV to IR evolution
in the $Adj$ and AT chiral gauge theories and compare with the conjectured
degree-of-freedom inequality (\ref{dfi}).


\subsection{UV Count}
\label{uvcount}

Given that we have required our theories to be asymptotically free, they are
weakly coupled in the UV, so we can identify the perturbative degrees of
freedom and calculate $f_{UV}$.  From the general formula (\ref{f}), we have
\beqs
f_{UV,R_{sc}}
& = & 2(N^2-1) + \frac{7}{4}\Big [ \frac{N(N+1)}{2} + (N+4)N \Big ]
\cr\cr 
& + & \frac{7}{8}\, p \, {\rm dim}(R_{sc}) \ ,
\label{fuv_rsc}
\eeqs
where the respective terms represent the contributions of the SU($N$) gauge
fields, the $S$ fermions, the $N+4$ copies of $\bar F$ fermions, and the
$R_{sc}$ fermions.  Explicitly, for the $Adj$ theory,
\beqs
f_{UV,Adj} & = & 2(N^2-1) + \frac{7}{4}\Big [ \frac{N(N+1)}{2} + (N+4)N \Big ]
\cr\cr & + & \frac{7}{8}\, p \, (N^2-1) 
\label{fuv_adjmodel}
\eeqs
and for the AT theory, with $N=2k$, 
\beqs
f_{UV,AT} &=& 2(N^2-1)+\frac{7}{4}\Big [ \frac{N(N+1)}{2} + (N+4)N \Big ]
\cr\cr & + & \frac{7}{8}\, p \, {N \choose N/2} \ . 
\label{fuv_kkmodel}
\eeqs
where ${a \choose b}$ is the binomial coefficient. 


\subsection{ $f_{IR}$ Calculations}

Next, we calculate $f_{IR}$ for the two types of chiral gauge theories
discussed above in the cases where the UV to IR evolution involves a high-scale
condensation in the respective channels (\ref{adjadjto1_adjmodel}) or
(\ref{kkchannel}), followed by sequential condensations in the
$S \times \bar F \to F$
channel.  Taking account of the $p(p+1)/2$ NGBs from the higher-scale symmetry
breaking at $\Lambda_{R_{sc}}$, we find, for either of these two types of
chiral gauge theories, for this type of infrared evolution below
$\Lambda_{R_{sc}}$,
\beqs
& & f_{IR,Adj;Adj \times Adj,S \times \bar F} = 
    f_{IR,AT;[k]_{2k} \times [k]_{2k},S \times \bar F}  \cr\cr
& \equiv & f_{IR,R_{sc}; R_{sc} \times R_{sc},S \times \bar F} \cr\cr
& = & 8N+1+\frac{7}{4}\Big [ \frac{N(N-1)}{2} + 4N \Big ] + \frac{p(p+1)}{2} 
\ , \cr\cr
& & 
\label{firbk_RR,bk}
\eeqs
where the subscript $R_{sc}$ identifies the chiral fermion representation in
the vectorlike subsector, the next subscript $R_{sc} \times R_{sc}$ is
shorthand for the MAC $R_{sc} \times R_{sc} \to 1$ in which the highest-scale
condensation takes place, and the last subscript, $S \times \bar F$ or $sym$
are shorthand for the two types of IR flow in the low-energy descendant theory,
namely sequential $S \times \bar F \to F$ condensation formation and gauge and
global symmetry breaking in the descendent theory, or confinement with
formation of massless composite fermions and retention of exact chiral symmetry
($sym$) in the infrared. Thus, the subscripts here and below placed after the
semicolon in quantities such as $f_{IR,Adj;Adj \times Adj,S \times \bar F}$ 
refer to the sequence of steps in the UV to IR evolution.

For the alternate type of evolution involving high-scale condensation in the 
respective channels (\ref{adjadjto1_adjmodel}) or
(\ref{kkchannel}), followed by confinement leading to massless gauge-singlet
composite fermions, we calculate, for either of our two types of chiral 
gauge theory with $R=R_{sc}$, 
\beqs & & f_{IR,Adj; Adj \times Adj, sym} = 
          f_{IR,AT; [k]_{2k} \times [k]_{2k},sym} \cr\cr
&\equiv & f_{IR,R_{sc}; R_{sc} \times R_{sc},sym} \cr\cr &
= & \frac{7}{4}\Big [ \frac{(N+4)(N+3)}{2} \Big ] + \frac{p(p+1)}{2} \ .
\label{firbk_RR,sym}
\eeqs
%


\subsection{Comparison with DFI for $Adj$ Theory} 

Using these inputs, we can now calculate $\Delta f$ for these chiral gauge
theories and compare with the conjectured degree-of-freedom inequality
(\ref{dfi}).  For both theories, if the UV to IR evolution is such as to lead
to a deconfined non-Abelian Coulomb phase, the perturbative degrees of freedom
are the same as in the UV, so the DFI is obeyed.  (The perturbative corrections
also obey the DFI \cite{dfvgt,dfcgt}.)  

We first discuss the possible cases for the theory with $R=Adj$.  If
$N$ and $p$ are such that the gauge interaction produces the 
high-scale condensation in the channel 
(\ref{xixicondensate_adjmodel}), followed by 
Eqs. (\ref{firstngb_rsc}) with (\ref{firbk_pzero}), we calculate 
\beqs
   & & (\Delta f)_{Adj;Adj \times Adj, S \times \bar F} \equiv 
f_{UV,Adj} - f_{IR,Adj;Adj \times Adj, S \times \bar F} \cr\cr
& = & \frac{1}{8}\Big [ 30N^2-50N-24+7pN^2-11p-4p^2 \big ] \ . \cr\cr
& & 
\label{Deltaf_adjadj_bk}
\eeqs
This is positive for $p$ satisfying the upper bound 
\beqs
& & p < \frac{1}{8}\Big [ 7N^2-11+\sqrt{49N^4+326N^2-800N-263} \ \Big ] \ . 
\cr\cr
& & 
\label{psol_adjmodel_RRbk}
\eeqs
The upper bound on the right-hand side of Eq. (\ref{psol_adjmodel_RRbk}) is
larger than the upper limit on $p$ imposed by the requirement of asymptotic
freedom, (\ref{p_upperbound_adjmodel}).  Hence, the conjectured
degree-of-freedom inequality (\ref{dfi}) is obeyed for all $N$ and allowed $p$
with this type of UV to IR evolution.

For the case where the low-energy effective $S\bar F$ theory confines without
any spontaneous chiral symmetry breaking, producing massless composite
fermions, we calculate
\beqs
   & & (\Delta f)_{Adj;Adj \times Adj,sym} \equiv 
f_{Adj,UV} - f_{IR,Adj;Adj \times Adj,sym} \cr\cr
& = & \frac{1}{8}\Big [ 30N^2 + 14N - 100 + 7pN^2-11p-4p^2 \Big ] \ . \cr\cr
& & 
\label{deltaf_adjmodel_RRsym}
\eeqs
This is positive for $p$ satisfying the upper bound 
\beqs
p & < &  \frac{1}{8}\Big [ 7N^2-11 \cr\cr
& + & \sqrt{49N^4+326N^2+224N-1479} \ \Big ] \ .
\label{psol_adjmodel_RRsym}
\eeqs
The upper bound on the right-hand side of Eq. (\ref{psol_adjmodel_RRsym}) is
larger than the upper limit on $p$ imposed by the requirement of asymptotic
freedom, (\ref{p_upperbound_adjmodel}).  Hence, the conjectured
degree-of-freedom inequality (\ref{dfi}) is also obeyed for all $N$ and allowed
$p$ with this type of UV to IR evolution.

As illustrative numerical examples, we may consider the cases $N=3$ and $N=4$.
In these cases, the respective upper bounds on $p$ from
Eq. ((\ref{p_upperbound_adjmodel}) are $p \le 3$, while the respective values
of the right-hand side of (\ref{psol_adjmodel_RRbk}) are 14.64 and 27.57 and
the respective values of the right-hand side of (\ref{psol_adjmodel_RRsym}) are
16.26 and 29.01.  Note that if $p$ is close to the upper bound $p_{b1z}$
arising from the requirement of asymptotic freedom, then $b_1$ is small, so
that $\alpha_{_{IR,2\ell}}$ is sufficiently small that the UV to IR evolution
is to a non-Abelian Coulomb phase, so that one knows that the DFI is satisfied
without going through the present analysis.

These expressions simplify in the limit $N \to \infty$ (with $p$ fixed) 
in Eq. (\ref{ln}).  We define rescaled degree-of-freedom 
measures that are finite in this limit, of the form 
\beq
\bar f \equiv \lim_{N \to \infty} \frac{f}{N^2} \ . 
\label{fbar}
\eeq
(We use the same notation, $\bar f$ for this $N \to \infty$ limit and for the
quantity (\ref{fbar_lnp}) defined in the LNP limit; the context will always 
make clear which limit is meant.)  We calculate 
\beq
\bar f_{UV,Adj} = \frac{37+7p}{8}  \ , 
\label{fuvbar_adjmodel_largeN}
\eeq
\beqs
& & \bar f_{IR,Adj; Adj \times Adj, S \times \bar F} = 
\bar f_{IR,Adj; Adj \times Adj, sym} = \frac{7}{8} \ , \cr\cr
& & 
\label{fir_adjmodel_largeN}
\eeqs
and hence 
\beqs
& & (\Delta \bar f)_{Adj; Adj \times Adj, S \times \bar F } = 
    (\Delta \bar f)_{Adj; Adj \times Adj, sym}
\cr\cr
& = & \frac{30+7p}{8}
\label{deltaf_adjmodel_largeN}
\eeqs
This obviously obeys the conjectured degree-of-freedom inequality (\ref{dfi}).


\subsection{Comparison with DFI for AT Theory} 

We next calculate $\Delta f$ for the AT chiral gauge theory with gauge group
$G={\rm SU}(N)$ with even $N=2k$ and $R_{sc}=[N/2]_N=[k]_{2k}$.  As noted
above, for values of $N$ and $p$ such that the UV to IR evolution is to a
deconfined non-Abelian Coulomb phase in the IR, the perturbative degrees of
freedom are the same as in the UV, and the conjectured degree-of-freedom
inequality is obeyed.

If $N$ and $p$ are such that the gauge interaction produces high-scale
condensation in the channel (\ref{adjadjto1_adjmodel}) followed at lower scales
by condensations in the successive $S \times \bar F \to F$ channels in SU($N$),
SU($N-1$), etc., then, using Eqs. (\ref{firstngb_rsc}) and (\ref{firbk_pzero}),
we compute
\beqs
& & (\Delta f)_{AT;[k]_{2k} \times [k]_{2k}, S \times \bar F} 
    \equiv f_{UV,AT}-f_{IR,AT;[k]_{2k} \times [k]_{2k},S \times \bar F}  \cr\cr
& = & \frac{1}{8}\Big [ 30N^2-50N-24+7p \, d_R -4p(p+1) \Big ] \ ,
\cr\cr
& & 
\label{deltaf_atmodel_RRbk}
\eeqs
where here $d_R \equiv {N \choose N/2}$.
This is positive for $p$ satisfying the upper bound 
\beqs
p & < & \frac{1}{8}\Big [ 7d_R - 4 \cr\cr
& + & \sqrt{480N^2-800N-368+49d_R^2-56d_R} \ \Big ] \ . \cr\cr
& & 
\label{psol_atmodel_RRbk}
\eeqs
The upper bound on the right-hand side of Eq. (\ref{psol_atmodel_RRbk}) is
larger than the upper limit on $p$ imposed by the requirement of asymptotic
freedom, (\ref{p_upperbound_atmodel}).  Hence, the conjectured
degree-of-freedom inequality (\ref{dfi}) is also obeyed for all $N$ and allowed
$p$ with this type of UV to IR evolution in the AT model.

For the alternate type of UV to IR evolution in which the 
low-energy effective $S\bar F$ theory confines without
any spontaneous chiral symmetry breaking, producing massless composite
fermions, we calculate
\beqs
& & (\Delta f)_{AT;[k]_{2k} \times [k]_{2k},sym} \equiv 
f_{AT,UV} - f_{IR,AT;[k]_{2k} \times [k]_{2k},sym} \cr\cr
& = & \frac{1}{8}\Big [ 30N^2+14N-100+7p \, d_R -4p(p+1) \Big ] \ . \cr\cr
& & 
\label{deltaf_atmodel_sym}
\eeqs
This is positive for $p$ satisfying the upper bound 
\beqs
p & < & \frac{1}{8}\Big [ 7d_R-4 \cr\cr
& + & \sqrt{480N^2+224N-1584+49d_R^2-56d_R} \ \Big ] \ . 
\cr\cr
& & 
\label{psol_atmodel_RRsym}
\eeqs
The upper bound on the right-hand side of Eq. (\ref{psol_atmodel_RRsym}) is
larger than the upper limit on $p$ imposed by the requirement of asymptotic
freedom, (\ref{p_upperbound_atmodel}).  Hence, the conjectured
degree-of-freedom inequality (\ref{dfi}) is also obeyed for all $N$ and allowed
$p$ with this type of UV to IR evolution in the AT model.

As numerical examples, for $N=4$ and $N=6$, the respective upper bounds on $p$
from Eq.  (\ref{p_upperbound_atmodel}) are $p \le 14$ and $p \le 7$, while the
respective right-hand sides of (\ref{psol_atmodel_RRbk}) are 14.05 and 38.86
and the respective right-hand sides of (\ref{psol_atmodel_RRsym}) are 16.22 and
40.56.  As before, it should be noted that if $p$ is close to the upper bound
from asymptotic freedom, $b_1$ is small, so that $\alpha_{_{IR,2\ell}}$ is
sufficiently small that the UV to IR evolution is to a non-Abelian Coulomb
phase, so that one knows that the conjecture degree-of-freedom inequality
(\ref{dfi}) is satisfied. 


\section{A Chiral Gauge Theory with $S \bar S$ Vectorlike Subsector} 
\label{ssbarmodel}


\subsection{Particle Content}
\label{ssbarmodel_content}

In this section we construct and study a chiral gauge theory with gauge group
SU($N$) and (massless) chiral fermion content such that the irreducibly chiral
part of the theory is the same as in our previous two theories, and the
vectorlike subsector consists of $p$ copies of fermions in $\{R + \bar R \}$
where $R$ is a non-self-conjugate higher-dimensional representation, namely the
symmetric rank-2 tensor, $S$.  Explicitly, the chiral fermions include
\begin{enumerate}

\item 
a symmetric rank-2 tensor representation, $S$, with corresponding field 
$\psi^{ab}_{i,L} = \psi^{ba}_{i,L}$, where $i=p+1$, 

\item 
$N+4$ copies of 
chiral fermions in the conjugate fundamental representation, $\bar F$, with 
fields $\chi_{a,j,L}$, where $j=1,...,N+4$, and 

\item 
$p$ copies of chiral fermions $\{S + \bar S\}$ in the symmetric rank-2 tensor 
and conjugate tensor representations, with fields  $\psi^{ab}_{i,L}$ and 
 $\psi_{ab,i,L}$, $i=1,...,p$. 

\end{enumerate}
This fermion content is summarized in Table
\ref{fermion_properties_ssbarmodel}. It is again clear that this theory is free
of any anomalies in gauged currents. We will refer to this as the $S \bar S$
theory.  


\subsection{Beta Function}
\label{ssbarmodel_betafunction}

The one- and two-loop terms in the beta function of this theory are 
\beq
(b_1)_{S\bar S} = 3N-2 -\frac{2p(N+2)}{3} 
\label{b1_ssbarmodel}
\eeq
and
\beqs
& & (b_2)_{S\bar S} = \frac{1}{2}(13N^2-30N+1+12N^{-1}) \cr\cr
& & - \frac{2p}{3}(8N^2+19N-12N^{-1}) \ . 
\label{b2_ssbarmodel}
\eeqs
The values of $p_{b1z,S\bar S}$ and $p_{b2z,S\bar S}$ are listed in Table 
\ref{pb12z_ssbarmodel}. 
As $p$ increases, the coefficient $(b_1)_{S \bar S}$ decreases and passes
through zero as $p$ ascends through the value
\beq
p_{b1z,S\bar S} = \frac{3(3N-2)}{2(N+2)} \ . 
\label{pb1z_ssbarmodel}
\eeq
The asymptotic freedom requirement requires $b_1 > 0$, i.e., 
\beq
p < \frac{3(3N-2)}{2(N+2)} \ . 
\label{p_upperbound_ssbarmodel}
\eeq
There are two marginal cases
to consider, consisting of values of $N$ and $p$ for which $(b_1)_{S \bar
S}=0$, so that one must determine the sign of $(b_2)_{S \bar S}$ to see if the
theory is asymptotically free.  These are the pairs $(N,p)=(6,3)$ and (22,4).
However, for both of these cases, $(b_2)_{S \bar S}$ is negative, so they are
excluded by the condition of asymptotic freedom. 
The upper bound $p_{b1z,S\bar S}$, is a monotonically increasing function of
$N$ for all physical $N$, increasing from 2.1 for $N=3$ and approaching the
limiting value 4.5 from below as $N \to \infty$.  The resultant physical,
integral values of $p$ that are allowed by the inequality
(\ref{p_upperbound_ssbarmodel}) are:
\beqs
& & p=0, \ 1, \ 2 \quad {\rm if} \ 3 \le N \le 6 \cr\cr
& & p=0, \ 1, \ 2, \ 3 \quad {\rm if} \ 7 \le N \le 22 \cr\cr
& & p=0, \ 1, \ 2, \ 3, \ 4 \quad {\rm if} \ N \ge 23 \ . 
\label{pvalues_ssbarmodel}
\eeqs
For small $p$ values, $(b_2)_{S \bar S}$ is positive, so the 
two-loop beta function $\beta_{\alpha,2\ell}$ has no IR zero.  The
coefficient $(b_2)_{S \bar S}$ decreases and passes through zero to negative
values as $p$ increases through the value 
\beq
(p_{b2z})_{S \bar S} = \frac{3(13N^3-30N^2+N+12)}{4(N+2)(8N^2+3N-6)} \ .
\label{pb2z_ssbarmodel}
\eeq
This is a monotonically increasing function of $N$ for all physical $N$,
increasing from the value $24/125=0.192$ at $N=3$ and approaching the limiting
value $39/32=1.21875$ from below as $N \to \infty$.  We list the values of
$p_{b1z,S\bar S}$ and $p_{b2z,S\bar S}$ in Table \ref{pb12z_ssbarmodel}.  Since
$p_{b1z,S\bar S} > p_{b2z,S \bar S}$, it follows that there is an interval
$(I_p)_{S \bar S}$ of values of $p$ for which $\beta_{\alpha,2\ell}$ has an IR
zero.  This zero occurs at $\alpha_{_{IR,2\ell,S\bar S}} = -4\pi (b_1)_{S \bar
S}/(b_2)_{S \bar S}$, where these coefficients were given above.  As $N \to
\infty$, the product $\alpha_{_{IR,2\ell,S\bar S}}N$ approaches the same limit
as for the $Adj$ model, given above in
Eq. (\ref{zetair_2loop_adjmodel_largeN}).


\subsection{Global Flavor Symmetry}
\label{global_flavor_symmetry_ssbar}

The classical global chiral flavor symmetry of this theory is 
\begin{widetext}
\beqs
G_{fl,cl,S\bar S} & = & {\rm U}(1+p)_S \otimes {\rm U}(N+4)_{\bar F} \otimes
 {\rm U}(p)_{\bar S} \cr\cr
& = & {\rm SU}(1+p)_S \otimes {\rm U}(1)_S \otimes 
{\rm SU}(N+4)_{\bar F} \otimes {\rm U}(1)_{\bar F} \otimes 
{\rm SU}(p)_{\bar S} \otimes {\rm U}(1)_{\bar S}  \ . 
\label{gflcl_ssbarmodel}
\eeqs
\end{widetext}
The representations of the various fermion fields under this symmetry are given
in Table \ref{fermion_properties_ssbarmodel}.  The corresponding global unitary
transformations are
\beq
\psi^{ab}_{i,L} \to \sum_{j=1}^{1+p} (U_S)_{ij} \, \psi^{ab}_{j,L} \ , 
\quad U_S \in {\rm U}(1+p)_S \ , 
\label{us_ssbarmodel}
\eeq
\beq
\chi_{a,i,L} \to \sum_{j=1}^{N+4}(U_{\bar F})_{ij} \, \chi_{a,j,L} \ , 
\quad U_{\bar F} \in {\rm U}(N+4)_{\bar F} \ , 
\label{ufbar_ssbarmodel}
\eeq
and 
\beq
\psi_{ab,i,L} \to \sum_{i=1}^p (U_{\bar S})_{ij} \, \psi_{ab,j,L} \ , 
\quad U_{\bar S} \in {\rm U}(p)_{\bar S} \ . 
\label{uxi_ssbarmodel}
\eeq

Each of the three global U(1) symmetries is broken by SU($N$) instantons.  As
before, we define the vector
\beqs
{\vec v} & = & \Big ( N_S T(S), \ N_{\bar F} T(\bar F), \ 
N_{\bar S} T(\bar S) \Big ) \cr\cr
 & = & \bigg [ (1+p)\Big (\frac{N+2}{2}\Big ), \ \frac{N+4}{2}, \ 
p\Big (\frac{N+2}{2}\Big ) \Big ] 
\ , 
\cr\cr
& & 
\label{anomvec_ssbarmodel}
\eeqs
where the basis is $(S,\bar F,\bar S)$, and we have used the values $N_S=1+p$,
$N_{\bar F}=N+4$, and $N_{\bar S}=p$.  In the same manner as before, we can
construct two linear combinations of the three original currents that are
conserved in the presence of SU($N$) instantons.  These have charges given by
\beq
{\vec Q}^{(j)} \equiv \Big ( Q^{(j)}_S, \ Q^{(j)}_{\bar F}, \ 
Q^{(j)}_{\bar S} \Big ), \quad j=1,2 \ , 
\label{qvec_ssbarmodel}
\eeq
where $j=1$ for U(1)$_1$ and $j=2$ for U(1)$_2$.  Next, we apply the 
conditions (\ref{u1inv}) and solve for the vectors of charges 
${\vec Q}^{(1)}$ and ${\vec Q}^{(2)}$ under the non-anomalous global symmetries
U(1)$_1$ and U(1)$_2$.  The condition (\ref{u1inv}) does not uniquely
determine the vectors ${\vec Q}^{(j)}$, $j=1, \ 2$.  We choose 
\beq
{\vec Q}^{(1)} = \Big ( N+4, \ -(1+p)(N+2), \ 0 \Big )
\label{qvec1_ssbarmodel}
\eeq
and
\beq
{\vec Q}^{(2)} = \Big ( 0, \ p(N+2), \ -(N+4) \Big ) \ . 
\label{qvec2_ssbarmodel}
\eeq
Then the (non-anomalous) global chiral flavor symmetry group of the theory is
\beq
G_{fl,S\bar S} = {\rm SU}(1+p)_{S} \otimes {\rm SU}(N+4)_{\bar F} \otimes 
{\rm SU}(p)_{\bar S} \otimes {\rm U}(1)_1 \otimes {\rm U}(1)_2 \ .
\label{gfl_ssbarmodel}
\eeq

For a given $N$ and $p$ that would lead to strong coupling in the infrared, we
check if the infrared theory could consist of confined, gauge-singlet massless
composite fermions that satisfy the 't Hooft anomaly-matching conditions.  The
possible gauge-singlet fermions that could, {\it a priori}, form are described
by the wavefunctions
\beqs
B_{ijk} & = & \bar F_{a,i,L} \, S^{ab}_{j,L} \, \bar F_{b,k,L}  \ , \ {\rm
  with}  \cr\cr 
& & 1 \le i,k \le N+4; \quad 1 \le j \le 1+p. \cr\cr
& & 
\label{bijell}
\eeqs
and
\beqs
B'_{ijk} & = & (\bar F^\dagger)^a_{i,L} \, \bar S_{ab,j,L} \, 
(\bar F^\dagger)^b_{k,L}  \ , \ {\rm with} \cr\cr
& & 1 \le i,k \le N+4, \quad 1 \le j \le p \ . \cr\cr
& & 
\label{bprimeijell}
\eeqs
The composite fermion $B_{ijk}$ transforms as a rank-2 antisymmetric tensor of
${\rm SU}(N+4)_{\bar F}$ and a fundamental representation of ${\rm
  SU}(1+p)_S$. From the analogue of the relation (\ref{qb}), its charges under
the two global abelian factor groups in 
$G_{fl,S\bar S}$, U(1)$_k$, $k=1,2$, are 
\beq
Q_B^{(1)} = -N-2p(N+2) 
\label{qb_u1_ssbarmodel}
\eeq
and
\beq
Q_B^{(2)} = 2p(N+2) \ . 
\label{qb_u2_ssbarmodel}
\eeq
The composite fermion $B'_{ijk}$ transforms as a rank-2 conjugate 
antisymmetric tensor of ${\rm SU}(N+4)_{\bar F}$ and a fundamental 
representation of ${\rm SU}(p)_{\bar S}$.  Its charges under
the two global abelian factor groups in 
$G_{fl,S\bar S}$, U(1)$_k$, $k=1,2$, are determined by the relation 
\beq
Q_{B'}^{(k)} = Q_{\bar S}^{(k)} - 2Q_{\bar F}^{(k)} \ , \ k=1,2. 
\label{qbprime}
\eeq
Hence, 
\beq
Q_{B'}^{(1)} = 2(1+p)(N+2) 
\label{qbprime_u1_ssbarmodel}
\eeq
and
\beq
Q_{B'}^{(2)} = -(N+4)-2p(N+2) \ . 
\label{qbprime_u2_ssbarmodel}
\eeq
We find that a hypothetical low-energy theory with these two massless composite
fermions would not satisfy the 't Hooft anomaly-matching conditions for any
nonzero value of $p$. (In the $p=0$ case, the theory degenerates to the $S\bar
F$ model, for which there is only the one type of composite fermion
(\ref{bij}), and the dynamics in the strongly coupled case would allow the
formation of this massless composite fermion.)  As with the $Adj$ and AT
theories, in the present $S \bar S$ theory, if $N$ and $p$ are such that the
theory becomes strongly coupled in the infrared, then the resultant fermion
condensation in the $S \times \bar S \to 1$ channel leaves, as the descendant
low-energy effective field theory below the scale of this condensation, the
$S\bar F$ theory. This is equivalent to the original $S\bar S$ theory with
$p=0$.


\subsection{UV to IR Evolution}

In order to investigate the nature of the UV to IR evolution in this $S \bar S$
theory, we first note that, again by design, the most attractive channel is 
\beq
S \times \bar S \to 1 \ , 
\label{ssbarto1_channel}
\eeq
preserving the SU($N$) gauge symmetry.  This has the attractiveness measure
\beq
\Delta C_2(S \times \bar S \to 1) = 2C_2(S) = \frac{2(N+2)(N-1)}{N} \ .
\label{deltac2_ssbar}
\eeq
That this is larger than the $\Delta C_2$ for the next-most-attractive channel
$S \times \bar F \to F$ is clear since
\beqs
& & \Delta C_2(S \times \bar S \to 1) - \Delta C_2(S \times \bar F \to F) 
\cr\cr
& = & C_2(S) > 0 \ . 
\label{ddeltassbar}
\eeqs
Using the rough estimate (\ref{alfcrit}), the minimal critical coupling for
condensation in the channel (\ref{ssbarto1_channel}) is
\beq
\alpha_{cr,S \times \bar S} = \frac{\pi N}{3(N+2)(N-1)} \ . 
\label{alfcrit_ssbar}
\eeq
In order to get an approximate measure of the size of the coupling
at the IR fixed point as compared with the minimum size for condensation, we
define the ratio
\beq
\rho_{_{IR,S \times \bar S}} \equiv \frac{\alpha_{IR,2\ell,S\bar S}}
{\alpha_{cr,S \times \bar S}} \ , 
\label{ratio_ssbar}
\eeq
depending on $N$ and $p \in (I_p)_{S \bar S}$. In Table
\ref{alfir_and_ratio_ssbarmodel} we list values of this ratio for a range of
$N$ and $p$ values.

As $N \to \infty$ (with $p$ fixed \cite{lnlnp}), the ratio $\rho_{_{IR,S \times
\bar S}}$ approaches the same limit as $\rho_{_{IR,Adj \times Adj}}$ in the
$Adj$ model, namely Eq. (\ref{alfalfcrit_ratio_adjmodel_largeN}), and the
specific values for the allowed range $p=4, \ 3, \ 2$ are the same as were
given in Eqs.
(\ref{alfalfcrit_ratio_adjmodel_largeN_peq4})-(\ref{alfalfcrit_ratio_adjmodel_largeN_peq2}). Also,
the same comments about the likely evolution to various IR phases that were
made in the $N \to \infty$ limit there also apply here.


\subsection{Comparison with DFI}

Since the theory is asymptotically free and hence weakly coupled in the UV, one
can enumerate the perturbative field degrees of freedom, with the result
\beqs
f_{UV,S \bar S M} & = & 2(N^2-1) + \frac{7}{4} \Big [ (2p+1)\frac{N(N+1)}{2}
  \cr\cr & + & (N+4)N \Big ] \ , 
\label{fuv_ssbarmodel}
\eeqs
where here and below, the subscript $S \bar S M$ means $S \bar S$ model.  For
values of $N$ and $p$ such that the beta function has an IR zero
$\alpha_{_{IR,2\ell}}$ at a value significantly smaller than $\alpha_{cr,S\bar
S}$, i.e., for which $\rho_{_{IR,S \bar S;S \times \bar S}}$ is well below
unity, one expects that the UV to IR evolution of this theory will not involve
any spontaneous chiral symmetry breaking but instead will lead to a deconfined
non-Abelian Coulomb phase in the infrared.  In this case, as for the other two
theories discussed above, at the weakly coupled perturbative level,
$f_{UV}=f_{IR}$, and the perturbative corrections obey the conjectured 
degree-of-freedom inequality (\ref{dfi}).

For values of $N$ and $p$ such that the beta function has no IR zero or an IR
zero $\alpha_{_{IR,2\ell}}$ that is moderately large, i.e., for which
$\rho_{_{IR,S \bar S; S \times \bar S}} \gsim O(1)$, one expects that as the
reference scale $\mu$ decreases sufficiently, the gauge coupling will become
large enough to produce a bilinear fermion condensate, and this condensate is
expected to be in the most attractive channel, (\ref{ssbarto1_channel}).  We
denote the scale at which this happens as $\Lambda_{S\bar S}$.  The associated
condensate is
\beq
\langle \psi^{ab \ T}_{i,L} C \psi_{ab,j,L} \rangle \ . 
\label{ssbar_condensate}
\eeq
A vacuum alignment argument suggests that the dynamics would yield condensates
with $i=j$, which would thus take the values $i=j=1,...,p$, namely 
\beq
\langle \sum_{i=1}^p \psi^{ab \ T}_{i,L} C \psi_{ab,j,L} \rangle \ . 
\label{ssbar_condensate_diag}
\eeq
The condensate (\ref{ssbar_condensate_diag}) preserves an SO($p$) isospin
symmetry defined by
\beqs
& & \psi^{ab}_{i,L} \to \sum_{i=1}^p {\cal O}_{ij} \psi^{ab}_{j,L} \ , \cr\cr
& & \psi_{ab,i,L} \to \sum_{i=1}^p {\cal O}_{ij} \psi_{ab,j,L}  \ . 
\label{soptran}
\eeqs
Here the orthogonal transformation ${\cal O} \in {\rm SO}(p)$ is in 1-1
correspondence with the special case of unitary transformation in ${\rm
SU}(p+1)_S$ that, furthermore, leaves the $i=p+1$ component of the
$(p+1)$-dimensional vector $(\psi^{ab}_{1,L},...,\psi^{ab}_{p+1,L})^T$
unchanged, and is also a special case of the unitary transformation in ${\rm
SU}(p)_{\bar S}$.  Assuming that the condensate takes the form
(\ref{ssbar_condensate_diag}), this process breaks the initial (non-anomalous)
global flavor symmetry group $G_{fl,S\bar S}$ to
\beq
G_{fl,S\bar S}' = {\rm SO}(p) \otimes {\rm SU}(N+4)_{\bar F}
 \otimes {\rm U}(1)' \ . 
\label{gfl_ssbarmodel_ssbarcond}
\eeq
Here the SU($N+4$) is (\ref{ufbar}), and the U(1)$'$ is the linear combination
of U(1)$_1$ and U(1)$_2$ for which the fields $(S,\bar F, \bar S)$ have charges
of the form $Q = (a,b,-a)$.  The $2p$ chiral fermions involved in the
condensate (\ref{ssbar_condensate}), namely the $S$ fields $\psi^{ab}_{i,L}$
and the $\bar S$ fields $\psi_{ab,i,L}$ with $i=1,...,p$, gain dynamical masses
of order $\Lambda_{S \bar S}$.  Note that this leaves the $(p+1)$'th component
$\psi^{ab}_{i,L}$ with $i=p+1$ still massless.  It follows that the number of
Nambu-Goldboson bosons produced by this spontaneous symmetry breaking of
$G_{fl,S \bar S}$ to $G_{fl,S \bar S}'$ is
\beq
o(G_{fl,S \bar S})-o(G_{fl,S \bar S}') = \frac{p(3p+5)}{2} \ . 
\label{ngb_ssbar_first}
\eeq

In the low-energy effective field theory applicable at scales $\mu <<
\Lambda_{S \bar S}$, one integrates out the now-massive $S$ and $\bar S$
fermions $\psi^{ab}_{i,L}$ and $\psi_{ab,i,L}$ with $i=1,...,p$. The resultant
global flavor symmetry group describing the massless degrees of freedom in this
low-energy effective theory is just that of the $S \bar F$ model, $G_{fl,S\bar
F}$ given in Eq. (\ref{gfle}). 

The further evolution of this $S \bar F$ theory into the infrared and the two
possibilities of confinement without chiral symmetry breaking or sequential
condensate formation in the $S \bar F \to F$ channel and associated gauge and 
global symmetry breaking have been reviewed above.  From these we can calculate
the resultant IR degrees of freedom and check the degree-of-freedom
inequality (\ref{dfi}).  

For the $S \times \bar S \to 1$ condensation followed by sequential $S \times
\bar F$ condensations in the SU($N$) theory, SU($N-1$) theory, etc., we have
\beqs
f_{IR,S\bar S M; S \times \bar S, S \times \bar F} & = & 
8N+1+\frac{7}{4}\Big [ \frac{N(N-1)}{2} + 4N \Big ] \cr\cr
& + & \frac{p}{2}(3p+5) \cr\cr
& = & \frac{N(7N+113)}{8} + 1 +\frac{p}{2}(3p+5) \ . 
\cr\cr
& & 
\label{fir_ssbarmodel_ssbar_sfbar}
\eeqs
Hence, 
\beqs
& & (\Delta f)_{S \bar S M; S \times \bar S, S \times 
\bar F} \equiv f_{UV,S \bar S M} -  f_{IR,S\bar S M; S \times \bar S,
S \times \bar F} \cr\cr
& = & \frac{1}{4}\Big [ 15N^2-25N-12 + p \{ 7N(N+1) - 2(3p+5) \} \Big ] \ .
\cr\cr
& & 
\label{deltaf_ssbarmodel_ssbar_sfbar}
\eeqs
This is positive for all $N$ and $p$ values of relevant here.  Explicitly, for
(nonnegative) $p$, $(\Delta f)_{S \bar S M; S \times \bar S, S \times \bar F}$
is positive if
\beqs
p & < & \frac{1}{12} \Big [ 7N(N+1) -10 \cr\cr
& + & \sqrt{49N^4+98N^3+269N^2-740N-188} \ \Big ] \ . \cr\cr
& & 
\label{p_upper_Deltaf_ssbarmodel_ssbar_sfbar}
\eeqs
The right-hand side of Eq. (\ref{p_upper_Deltaf_ssbarmodel_ssbar_sfbar}) is
greater than the upper bound $p_{b1z}$ allowed by asymptotic freedom.  For
example, for $N=3$, the physical, integral values of $p$ are required by
asymptotic freedom to be $\le 2$, whereas the right-hand side of
(\ref{p_upper_Deltaf_ssbarmodel_ssbar_sfbar}) is 12.95; and for $N=4$,
asymptotic freedom again requires $p \le 2$, whereas the right-hand side of
(\ref{p_upper_Deltaf_ssbarmodel_ssbar_sfbar}) is 22.61, and similarly for
larger values of $N$.

For a UV to IR evolution involving $S \bar S \to 1$ condensation followed by 
confinement without spontaneous chiral symmetry breaking, we find 
\beqs
f_{IR,S\bar S M; S \times \bar S, sym} 
& = & \frac{7}{4}\Big [ \frac{(N+4)(N+3)}{2}
\Big ] + \frac{p(3p+5)}{2} \cr\cr
& & 
\label{fir_ssbarmodel_ssbar_sym}
\eeqs
and hence 
\beqs
& & (\Delta f)_{S \bar S M; S \times \bar S, sym} \equiv f_{UV,S \bar S M} - 
f_{IR,S\bar S M; S \times \bar S, sym} \cr\cr 
& = & \frac{1}{4}\Big [ 15N^2+7N-50 + p\{ 7N(N+1)-2(3p+5)\} \Big ] \ . 
\cr\cr
& & 
\label{deltaf_ssbarmodel_ssbar_sym}
\eeqs
This is positive for all $N$ and $p$ values of relevant here.  Explicitly, for
(nonnegative) $p$, $(\Delta f)_{S \bar S M; S \bar S, S \bar F}$ is positive if
\beqs
p & < & \frac{1}{12} \Big [ 7N(N+1) -12 \cr\cr
& + & \sqrt{49N^4+98N^3+269N^2+28N-1100} \ \Big ] \ . \cr\cr
& & 
\label{p_upper_Deltaf_ssbarmodel_ssbar_sym}
\eeqs
The right-hand side of Eq. (\ref{p_upper_Deltaf_ssbarmodel_ssbar_sym}) is
greater than the upper bound $p_{b1z}$ allowed by asymptotic freedom.  For
example, for $N=3$, $p \le 2$ for asymptotic freedom, while the right-hand side
of (\ref{p_upper_Deltaf_ssbarmodel_ssbar_sym}) is 13.63; and for $N=4$, again,
$p \le 2$ for asymptotic freedom, while the right-hand side of
(\ref{p_upper_Deltaf_ssbarmodel_ssbar_sym}) is 23.23, and similarly for larger
values of $N$.

In the $N \to \infty$ limit (\ref{ln}) (with $p$ fixed), we have
\beq
\bar f_{UV,S\bar S} = \frac{37+14p}{8} 
\label{fuvbar_ssbarmodel}
\eeq
and
\beq
\bar f_{IR,S\bar S M; S \times \bar S, S \times \bar F} = 
\bar f_{IR,S\bar S M; S \times \bar S, sym} = \frac{7}{8} \ , 
\label{firbar_ssbarmodel}
\eeq
and hence
\beqs
& & (\Delta \bar f)_{S \bar S M; S \times \bar S, S \times \bar F} \equiv 
\bar f_{UV,S \bar S M} - \bar f_{UV,S \bar S M; S \times \bar S, S 
\times \bar F } 
\cr\cr
& = & (\Delta \bar f)_{S \bar S M; S \times \bar S, sym} \equiv 
\bar f_{UV,S \bar S M} - \bar f_{UV,S \bar S M; S \times \bar S, sym} 
\cr\cr
& = & \frac{15+7p}{4} \ . 
\label{deltafbar_ssbmodel}
\eeqs
This difference is manifestly positive, in agreement with the conjectured 
degree-of-freedom inequality (\ref{dfi}). 


\section{Conclusions}
\label{conclusions}

In summary, we have constructed three asymptotically free chiral gauge theories
and analyzed their renormalization-group evolution from the ultraviolet to the
infrared.  These theories have the gauge group SU($N$) and massless fermions
transforming according to a symmetric rank-2 tensor representation, $S$, and
$N+4$ copies of a conjugate fundamental representation, $\bar F$, together with
a vectorlike subsector with $p$ copies of fermions in higher-dimensional
representation(s). We first studied two theories with the vectorlike fermions
in different self-conjugate representations, namely theories with $p$ copies of
fermions in (a) the adjoint representation and (b) in the antisymmetric
rank-$k$ tensor representation of ${\rm SU}(2k)$.  We have also studied a third
type of theory, with a vectorlike subsector consisting of $p$ pairs of fermions
transforming as $\{S + \bar S\}$.  We have presented results on beta functions,
IR zeros of these beta functions, and possible types of UV to IR evolution.  In
analyzing fermion condensate formation, we have made use of the
most-attractive-channel approach. We have shown that for these three types of
chiral gauge theories, the various types of likely UV to IR evolution satisfy
the conjectured degree-of-freedom inequality (\ref{dfi}) for all relevant
values of $N$ and $p$. It is hoped that the new chiral gauge theories
constructed and analyzed here may serve as useful theoretical laboratories for
the study of chiral gauge theories in future work. 


\begin{acknowledgments}
This research was partially supported by the NSF grant NSF-PHY-13-16617.
\end{acknowledgments}


\begin{appendix}

\section{Beta Function Coefficients and Relevant Group Invariants}
\label{bn}

For reference, we list the one-loop and two-loop coefficients
\cite{b1,b2} in the beta function (\ref{beta}) for a non-Abelian
chiral gauge theory with gauge group $G$ and a set of chiral fermions
comprised of $N_i$ fermions transforming according to the representations 
$\{ R_i \}$. 
\beq
b_1 = \frac{1}{3}\Big [ 11 C_2(G) - 2 \sum_{R_i}N_i T(R_i) \Big ]
\label{b1}
\eeq
and
\beq
b_2=\frac{1}{3}\Big [ 34 C_2(G)^2 - 
2\sum_{R_i} N_i \{ 5C_2(G)+3C_2(R_i)\} T(R_i) \Big ] \ . 
\label{b2}
\eeq

We list below the group invariants that we use for the relevant case 
$G={\rm SU}(N)$.  We have $C_2(G)=C_2(Adj)=T(Adj)=N$, and, as in the text, 
we use the symbols $F$ for $\fund$ and $S$ for $\sym$.  We have 
\beq
C_2(F) = \frac{N^2-1}{2N} \ , \quad T(F) = \frac{1}{2} \ , 
\label{casimir_fund}
\eeq
\beq
C_2(S) = \frac{(N+2)(N-1)}{N} \ , \quad T(S) = \frac{N+2}{2} \ , 
\label{casimir_sym}
\eeq
\beq
C_2([k]_N) = \frac{k(N+1)(N-k)}{2N} \ , 
\label{c2kn}
\eeq
and
\beq
T([k]_N) = \frac{1}{2}{N-2 \choose k-1} \ .
\label{tkn}
\eeq
Hence, for our case $N=2k$, 
\beq
C_2([k]_{2k}) = \frac{k(2k+1)}{4}
\label{c2k2k}
\eeq
and
\beq
T([k]_{2k}) = \frac{(2k-2)!}{2[(k-1)!]^2} \ . 
\label{tk2k}
\eeq
%


\end{appendix} 



\newpage

\begin{table}
\caption{\footnotesize{Properties of fermions in the chiral gauge theories with
vectorlike subsector consisting of $p$ copies of fermions in the self-conjugate
representation $R=R_{sc}$. The entries in the columns are: (i) fermion, (ii)
representation of the SU($N$) gauge group, (iii) number of copies, and
representations (charges for abelian factors) of the respective factor groups
in the global flavor symmetry group: (iv) ${\rm SU}(N+4)_{\bar F}$, (v) ${\rm
SU}(p)_{R_{sc}}$, (vi) U(1)$_1$, (vii) U(1)$_2$.  The notation for the fermion
$\xi$ in the $R_{sc}$ is generic; specifically, this is $\xi^a_{b,i,L}$ for the
$Adj$ model and $\xi^{a_1,...,a_k}_{i,L}$ for the AT model (with $N=2k$).  See
text for further discussion.}}
\begin{center}
\begin{tabular}{|c|c|c|c|c|c|c|} 
\hline\hline
fermion  & SU($N$) & no. copies & ${\rm SU}(N+4)_{\bar F}$ & 
${\rm SU}(p)_{R_{sc}}$ & U(1)$_1$ & U(1)$_2$ \\
\hline 
$S:\ \psi^{ab}_L$ & $\sym$ & 1 &    1    &   1   &   $N+4$ &  $2pT_{R_{sc}}$ \\
$\bar F: \ \chi_{a,i,L}$ & $\overline{\fund}$ & $N+4$ & $\fund$ & 1 &
$-(N+2)$ & 0 \\
$R_{sc}$: \ $\xi_L$ & $R_{sc}$ & $p$ & 1 & $\fund$ & 0 & $-(N+2)$ \\
\hline\hline
\end{tabular}
\end{center}
\label{fermion_properties_rscmodels}
\end{table}


\begin{table}
\caption{\footnotesize{Values of $p_{b1z,Adj}$ and $p_{b2z,Adj}$ in the $Adj$
    theory as functions of $N$.}}
\begin{center}
\begin{tabular}{|c|c|c|} 
\hline\hline
$N$ & $p_{b2z,Adj}$ & $p_{b1z,Adj}$
\\ \hline
3      & 0.3333   & 3.5000  \\
4      & 0.5391   & 3.7500  \\
5      & 0.6690   & 3.9000  \\
6      & 0.7578   & 4.0000  \\
7      & 0.8222   & 4.0714  \\
8      & 0.8708   & 4.1250  \\
9      & 0.90895  & 4.1667  \\
10     & 0.9396   & 4.2000  \\
11     & 0.9647   & 4.2773  \\ 
12     & 0.9857   & 4.2500  \\
13     & 1.0035   & 4.2692  \\
14     & 1.0187   & 4.2857  \\
15     & 1.0320   & 4.3000  \\
$10^2$ & 1.1906   & 4.4700  \\
$10^3$ & 1.2159   & 4.4970  \\
$\infty$ & 1.21875& 4.5000  \\
\hline\hline
\end{tabular}
\end{center}
\label{pb12z_adjmodel}
\end{table}
%


\begin{table}
\caption{\footnotesize{Values of $\alpha_{_{IR,2\ell,Adj}}$ and $\rho_{_{IR,Adj
\times Adj}}$ in the $Adj$ theory for an illustrative range of values of $N$
and, for each $N$, the values of $p$ in the respective interval
$(I_p)_{Adj}$.}}
\begin{center}
\begin{tabular}{|c|c|c|c|} 
\hline\hline
$N$ & $p$ & $\alpha_{_{IR,2\ell,Adj}}$ & $\rho_{_{IR,Adj \times Adj}}$ 
\\ \hline
3  &  1  &  1.96   &  5.63  \\
3  &  2  &  0.471  &  1.35  \\
3  &  3  &  0.0982 &  0.281 \\
\hline
4  &  1  &  2.34   &  8.95  \\
4  &  2  &  0.470  &  1.80  \\
4  &  3  &  0.120  &  0.457 \\
\hline
5  &  1  &  2.75   & 13.1   \\
5  &  2  &  0.448  & 2.14   \\
5  &  3  &  0.121  & 0.579  \\
\hline
6  &  1  &  3.24   & 18.6  \\
6  &  2  &  0.4215 & 2.415 \\
6  &  3  &  0.117  & 0.669 \\
\hline
7  &  1  &  3.88   & 25.9  \\
7  &  2  &  0.395  & 2.64  \\
7  &  3  &  0.110  & 0.738 \\
7  &  4  &  0.00504& 0.0337 \\
\hline 
8  &  1  &  4.75   & 36.3   \\
8  &  2  &  0.370  & 2.82   \\
8  &  3  &  0.104  & 0.793  \\
8  &  4  &  0.00784& 0.0599 \\
\hline
13 &  2  &  0.275   & 3.42   \\
13 &  3  &  0.0768  & 0.954  \\
13 &  4  &  0.0109  & 0.135  \\
\hline
14 &  2  &  0.261   & 3.49   \\
14 &  3  &  0.0728  & 0.973  \\
14 &  4  &  0.01075 & 0.144  \\
\hline 
15 &  2  &  0.249   & 3.56   \\
15 &  3  &  0.0692  & 0.991  \\
15 &  4  &  0.0106  & 0.152  \\
\hline\hline
\end{tabular}
\end{center}
\label{alfir_and_ratio_adjmodel}
\end{table}
%


\begin{table}
\caption{\footnotesize{Values of $p_{b1z,AT}$, $p_{b2z,AT}$, $p_{max}$, and the
    intervals $(I_p)_{AT}$ as functions of
$N$ in the AT model with gauge group SU($N$) with $N=2k$.}}
\begin{center}
\begin{tabular}{|c|c|c|c|c|}
\hline\hline
$N$ & $p_{b2z,AT}$ & $p_{b1z,AT}$ & $p_{max}$ & $(I_p)_{AT}$ 
\\ \hline
4      & 2.509   & 15     &  14  &  $3 \le p \le 14$ \\ 
6      & 1.590   & 8      &   7  &  $2 \le p \le 7$  \\
8      & 0.665   & 3.3    &   3  &  $1 \le p \le 3$  \\
10     & 0.235   & 1.2    &   1  &  $p=1$            \\
\hline\hline
\end{tabular}
\end{center}
\label{pb12z_atmodel}
\end{table}
%



\begin{table}
\caption{\footnotesize{Values of $\alpha_{_{IR,2\ell,AT}}$ and
$\rho_{_{IR,AT}}$ in the AT theory for the relevant values of $N$ and, for each
$N$, the values of $p$ in the respective interval $(I_p)_{AT}$.}}
\begin{center}
\begin{tabular}{|c|c|c|c|} 
\hline\hline
$N$ & $p$ & $\alpha_{_{IR,2\ell,AT}}$ & $\rho_{_{IR,AT}}$ 
\\ \hline
4  &  3  & 11.170   & 26.67   \\
4  &  4  &  3.371   &  8.05   \\
4  &  5  &  1.8345  &  4.38   \\
4  &  6  &  1.178   &  2.81   \\
4  &  7  &  0.814   &  1.94   \\
4  &  8  &  0.583   &  1.39   \\
4  &  9  &  0.422   &  1.01   \\
4  & 10  &  0.305   &  0.728  \\
4  & 11  &  0.215   &  0.514  \\
4  & 12  &  0.144   &  0.345  \\
4  & 13  &  0.0871  &  0.208  \\
4  & 14  &  0.0398  &  0.095  \\
\hline
6  &  2  &  4.021  & 20.16    \\
6  &  3  &  0.974  &  4.88    \\
6  &  4  &  0.460  &  2.29    \\
6  &  5  &  0.242  &  1.21    \\
6  &  6  &  0.125  &  0.625   \\
6  &  7  &  0.0508 &  0.255   \\
\hline
8  &  1  &  1.290  & 11.08    \\
8  &  2  &  0.183  &  1.57    \\
8  &  3  &  0.0241 &  0.207   \\
\hline
10 &  1  &  0.0360  & 0.473   \\
\hline\hline
\end{tabular}
\end{center}
\label{alfir_and_ratio_atmodel}
\end{table}
%



\begin{table}
\caption{\footnotesize{Properties of fermions in the $S \bar S$ theory with
vectorlike subsector consisting of $p$ copies of fermions in the 
$\{S + \bar S\}$
representations. The entries in the columns are: (i) fermion, (ii)
representation of the SU($N$) gauge group, (iii) number of copies, and
representations (charges for abelian factors) of the respective factor groups
in the global flavor symmetry group $G_{fl,S \bar S}$: (iv) ${\rm SU}(1+p)_S$;
(v) ${\rm SU}(N+4)_{\bar F}$, (vi) ${\rm SU}(p)_{\bar S}$, (vi) U(1)$_1$, (vii)
U(1)$_2$. See text for further discussion.}}
\begin{center}
\begin{tabular}{|c|c|c|c|c|c|c|c|} 
\hline\hline
fermion  & SU($N$) & no. copies & ${\rm SU}(1+p)_S$ & ${\rm SU}(N+4)_{\bar F}$
 & ${\rm SU}(p)_{\bar S}$ & U(1)$_1$ & U(1)$_2$ \\
\hline 
$S:\ \psi^{ab,i}_L$ & $\sym$ & $1+p$ & $\fund$  & 1 & 1 & $N+4$ & 0 \\
$\bar F: \ \chi_{a,j,L}$ & $\overline{\fund}$ & $N+4$ & 1 & $\fund$ & 1 &
$-(1+p)(N+2)$ & $p(N+2)$ \\
$\bar S:\ \psi_{ab,k,L}$ & $\overline{\sym}$ & $p$ & 1 & 1 & $\fund$ & 0 & 
$-(N+4)$ \\
\hline\hline
\end{tabular}
\end{center}
\label{fermion_properties_ssbarmodel}
\end{table}


\begin{table}
\caption{\footnotesize{Values of $p_{b1z,S\bar S}$ and $p_{b2z,S\bar S}$ in the
    $S \bar S$ theory, as functions of $N$.}}
\begin{center}
\begin{tabular}{|c|c|c|} 
\hline\hline
$N$ & $p_{b2z,S \bar S}$ & $p_{b1z,S \bar S}$ 
\\ \hline
3      & 0.1920   & 2.1000  \\
4      & 0.3433   & 2.5000  \\
5      & 0.4573   & 2.7857  \\
6      & 0.5456   & 3.0000  \\
7      & 0.6159   & 3.1667  \\
8      & 0.6730   & 3.3000  \\
9      & 0.7203   & 3.4091  \\
10     & 0.7602   & 3.5000  \\
20     & 0.9642   & 3.9455  \\
25     & 1.0106   & 4.0555  \\
50     & 1.1098   & 4.2692  \\
$10^2$ & 1.1630   & 4.3824  \\
$10^3$ & 1.2131   & 4.4880  \\
$\infty$ & 1.21875& 4.5000  \\
\hline\hline
\end{tabular}
\end{center}
\label{pb12z_ssbarmodel}
\end{table}
%


\begin{table}
\caption{\footnotesize{Values of $\alpha_{_{IR,2\ell,S\bar S}}$ and
$\rho_{_{IR,S \times \bar S}}$ for $3 \le N \le 8$ in the $S \bar S$ theory,
and, for each $N$, the values of $p \in (I_p)_{S \bar S}$.}}
\begin{center}
\begin{tabular}{|c|c|c|c|} 
\hline\hline
$N$ & $p$ & $\alpha_{_{IR,2\ell,S\bar S}}$ & $\rho_{_{IR,S \times \bar S}}$ 
\\ \hline
3  &  1  &  0.684   & 2.18   \\
3  &  2  &  0.0278  & 0.0885 \\
\hline
4  &  1  &  0.857   & 3.68   \\
4  &  2  &  0.113   & 0.486  \\
\hline
5  &  1  &  0.989   & 5.29   \\
5  &  2  &  0.153   & 0.819  \\
\hline
6  &  1  &  1.106   & 7.04  \\
6  &  2  &  0.173   & 1.10  \\
\hline
7  &  1  &  1.219   & 8.98  \\
7  &  2  &  0.182   & 1.34  \\
7  &  3  &  0.015   & 0.111 \\
\hline 
8  &  1  &  1.334   &11.15 \\
8  &  2  &  0.186   & 1.55 \\
8  &  3  &  0.0245  & 0.204 \\
\hline\hline
\end{tabular}
\end{center}
\label{alfir_and_ratio_ssbarmodel}
\end{table}
%



\begin{thebibliography}{99}
\bibitem{casimir}
%
The Casimir invariants $C_2(R)$ and $T(R)$ are defined as 
$\sum_a \sum_j {\cal D}_R(T_a)_{ij} {\cal D}_R(T_a)_{jk}=C_2(R)\delta_{ik}$ 
and 
$\sum_{i,j} {\cal D}_R(T_a)_{ij} {\cal D}_R(T_b)_{ji}=T(R) \delta_{ab}$, 
where $R$ is the
representation, $T_a$ are the generators of $G$, and $D_R$ is the matrix
representation ({\it Darstellung}) of $R$.
\bibitem{acon}
%
The coefficient $2$ multiplying $N_v$ in Eq. (\ref{f}) has a clear
physical meaning, as arising from the two dynamical transverse degrees of
freedom in the massless vector boson field. As discussed in \cite{cgt}, this is
to be contrasted with a different effective measure of field degrees of freedom
which uses the term $62N_v$ rather than $2N_v$.
\bibitem{sc}
%
The term $(7/4)N_f$ for non-self-conjugate fermions is the sum of a term 
$(7/8)N_f$ for the fermions and $(7/8)N_f$ for the antifermions; in the case of
a self-conjugate, Majorana fermion, these are the same.
\bibitem{lnlnp}
%
Note that for all of the three new chiral gauge theories that we construct and
study in this paper, namely the $Adj$, AT, and $S \bar S$ theories, when we
take a large-$N$ limit, we keep $p$ fixed rather than having it increase
like $rN$.  The reason for this difference relative to the $Sp$ theory is that
since each of these three new theories involves $p$ fermions in two-index
representations of SU($N$), it follows that for these fermions, $T(R) \propto
N$ for large $N$, in contrast to the $Sp$ model, where the $p$ fermions are in
the $F$ or $\bar F$ representation, with $T(F)=1/2$ independent of $N$.  Hence,
to keep $b_1$ positive as required for asymptotic freedom, the large-$N$ limit
for these three theories is taken with $p$ fixed.
\bibitem{cgtmp}
%
Note that in the line below Eq. (5.3) in \cite{cgt}, there is a misprint; the
text should read ``satisfying the inequality for $r \le \sqrt{15/14}$'', as is
obvious from Eq. (4.9) in \cite{cgt}.
\bibitem{anom}
%
The anomaly produced by chiral fermions transforming according to the 
representation $R$ of a group $G$ is defined as ${\rm Tr}_R(T_a,\{T_b,T_c\}) =
Anom(R)d_{abc}$, where the $d_{abc}$ are the totally symmetric structure 
constants of the corresponding Lie algebra.
\bibitem{srgt}
%
As noted in the text, we confine ourselves here to chiral gauge theories
without fundamental scalar fields; there have also been many studies of
renormalization-group behavior of chiral gauge theories with both fermion and 
scalar fields.
\bibitem{npreal}
%
Here and elsewhere, when an expression is given for $N$ and/or $p$
that formally evaluates to a non-integral real value, it is understood
implicitly that one infers an appropriate integral value from it, either the
greatest integer smaller than the real value, or the least integer greater than
the real value, or the closest integer, depending on the context.
\bibitem{vacalign}
%
We recall that an analogous vacuum alignment argument is used in quantum
chromodynamics to infer that among the various $3 \times \bar 3 \to 1$ quark
condensates, $\langle \bar u u \rangle$, $\langle \bar d d \rangle$, and 
$\langle \bar s s \rangle$ form, but others such as $\langle \bar u d \rangle$
and $\langle \bar u s \rangle$, and $\langle \bar d s \rangle$ do not.
\bibitem{adjanom}
%
Parenthetically, we note that in the $Adj$ theory, the anomaly in the global
U(1)$_{R_{sc}}$ symmetry is analogous to the anomaly in the U(1)$_R$ global
symmetry in supersymmetric SU($N$) gauge theory.
\end{thebibliography}
\end{document}